\documentclass[acmsmall,screen]{acmart}

\usepackage{amsmath,amssymb,amsfonts}
\usepackage{array,framed}
\usepackage{balance}
\usepackage{booktabs}
\usepackage{bm} 
\usepackage{csvsimple}
\usepackage[table, dvipsnames]{xcolor}
\usepackage{xcolor}
\usepackage{colortbl}
\usepackage{enumerate}
\usepackage{enumitem}
\usepackage[T1]{fontenc} 
\usepackage{setspace}
\usepackage[bookmarksnumbered,unicode]{hyperref}
\usepackage{latexsym,fancyhdr}
\usepackage{listings}
\usepackage{mathtools}
\usepackage{makecell}
\usepackage{totcount}
\usepackage{multirow}
\usepackage{nicefrac}
\usepackage{pythonhighlight}
\usepackage{setspace}
\usepackage{siunitx}
\usepackage{soul}
\usepackage{setspace}
\usepackage{tabularx}
\usepackage{textcomp}
\usepackage{totcount}
\usepackage{tikz}
\usepackage{url}
\usepackage{xparse}
\usepackage{xspace}
\usepackage{color}
\usepackage{
  float,
  epsfig,
  wrapfig,
  graphics,
  graphicx,
  subcaption
}
\usepackage{tcolorbox}
\usepackage{nicematrix}

\usepackage{natbib}
\usepackage{microtype}
\usepackage{tabularray}
\usepackage{etoolbox}
\usepackage{comment}
\usepackage[T1]{fontenc}
\usepackage{tgcursor}
\usepackage{longtable}
\usepackage{wasysym}
\usepackage{bm}
\usepackage{minibox}
\usepackage{minted}
\usepackage{pgfkeys}
\usepackage{refcount}
\usepackage{xurl}
\usepackage{mdframed}
\usepackage{algorithm}
\usepackage{algorithmicx}
\usepackage{algcompatible}
\usepackage{algpseudocode}
\usepackage{fix-cm}

\newif \ifauthorcomment
\authorcommenttrue

\newtotcounter{comments}

\newcommand{\ptitle}[1]{
\noindent{\bf \IfEndWith{#1}{.}{#1}{#1.}}
}
\newcommand{\PP}[1]{
\noindent{\bf {#1.}}}

\newcommand{\ie}{\textit{i}.\textit{e}., }
\newcommand{\eg}{\textit{e}.\textit{g}., }

\definecolor{Gray}{gray}{0.5}
\definecolor{RED}{rgb}{1,0.64,0.68}
\definecolor{GREEN}{rgb}{0.48,0.81,0.48}
\definecolor{codegreen}{rgb}{0,0.6,0}
\definecolor{codegray}{rgb}{0.5,0.5,0.5}
\definecolor{codepurple}{rgb}{0.58,0,0.82}
\definecolor{backcolour}{rgb}{0.95,0.95,0.92}
\definecolor{backwhite}{rgb}{1,1,1}
\definecolor{backgray}{rgb}{0.95,0.95,0.95}
\definecolor{highlightcolor}{gray}{0.2}

\setlist[itemize]{leftmargin=.15in, topsep={2pt}, partopsep={0pt}}

\usetikzlibrary{
  shapes.geometric,
  arrows,
  external,
  pgfplots.groupplots,
  matrix
}

\newcommand{\reducedstrut}{\vrule width 0pt height 1.1\ht\strutbox depth 0.8\dp\strutbox\relax}

\newtcolorbox{boxI}{
    colback = lightgray!10, 
    colframe = black, 
    boxrule = 1pt, 
    toprule = 1pt, 
    arc = 2pt,
    left = 1pt,
    right = 1pt,
    bottom = 1pt,
    top = 1pt
}

\newcounter{observcntr}
\newcommand*{\observ}[1]{%
    \stepcounter{observcntr}%
    \begin{center}
    \vspace{-10px}
        \begin{boxI}
        \textbf{Takeaway~\arabic{observcntr}: }{#1}.
        \end{boxI}
    \vspace{-5px}
    \end{center}
}

\definecolor{codegray}{RGB}{240, 240, 240}
{\endMakeFramed}

\usetikzlibrary{
  arrows.meta,
  tikzmark,
  calc,
  positioning
}
\usetikzmarklibrary{listings}

\newcommand{\textgraybackcolor}[1]{
    \begingroup\fboxsep=-0.5pt\colorbox{Gray!15}{\reducedstrut \textbf{#1}}\endgroup\looseness=-1
}

\newcommand\rurl[1]{%
  \href{http://#1}{\nolinkurl{#1}}%
}
\setlist{leftmargin=3.5mm}

\definecolor{dkgreen}{rgb}{0,0.6,0}
\definecolor{gray}{rgb}{0.5,0.5,0.5}
\definecolor{mauve}{rgb}{0.58,0,0.82}

\AtBeginDocument{%
  }

\author{Kiho Lee}
\affiliation{%
  \institution{ETRI}
  \country{Republic of Korea}} 
\email{kiho@etri.re.kr}

\author{Jungkon Kim}
\affiliation{%
  \institution{Samsung Research}
  \country{Republic of Korea}}
\email{jungkon117@g.skku.edu}

\author{Doowon Kim}
\affiliation{%
  \institution{University of Tennessee, Knoxville}
  \country{Republic of Korea}}
\email{doowon@utk.edu}

\author{Hyoungshick Kim}
\affiliation{%
  \institution{Sungkyunkwan University}
  \country{Republic of Korea}}
\email{hyoung@skku.edu}

\begin{document}

\title{A Systematic Evaluation of Parameter-Efficient Fine-Tuning Methods for the Security of Code LLMs}



\begin{abstract}
Code-generating Large Language Models (LLMs) significantly accelerate software development. However, their frequent generation of insecure code presents serious risks. We present a comprehensive evaluation of seven parameter-efficient fine-tuning (PEFT) techniques, demonstrating substantial gains in secure code generation without compromising functionality. Our research identifies prompt-tuning as the most effective PEFT method, achieving an 80.86\% \texttt{Overall-Secure-Rate} on CodeGen2 16B, a 13.5-point improvement over the 67.28\% baseline. Optimizing decoding strategies through sampling temperature further elevated security to 87.65\%. This equates to a reduction of approximately 203,700 vulnerable code snippets per million generated. Moreover, prompt and prefix tuning increase robustness against poisoning attacks in our TrojanPuzzle evaluation, with strong performance against CWE-79 and CWE-502 attack vectors. Our findings generalize across Python and Java, confirming prompt-tuning's consistent effectiveness. This study provides essential insights and practical guidance for building more resilient software systems with LLMs.

\end{abstract}

\maketitle

\section{Introduction}~\label{sec:intro}

LLM-based coding assistants have significantly enhanced developer productivity by providing intelligent code suggestions~\cite{Staff_2024,Paul_2025,allthingsopen2025,unpacking2025}. However, this advancement introduces substantial security risks: nearly half (48\%) of AI-generated code contains vulnerabilities ranging from SQL injection to hard-coded credentials~\cite{darkreading2025ai,pearce2022asleep}. When deployed into production, these vulnerabilities expand the attack surface, significantly increasing risk for end-users and potentially outweighing productivity gains.

The threat landscape extends beyond inadvertent vulnerabilities. Coordinated poisoning attacks demonstrate how code-generation LLMs can become sophisticated vectors of malicious activity~\cite{schuster2021you,wan2022you,aghakhani2023trojanpuzzle}. TrojanPuzzle~\cite{aghakhani2023trojanpuzzle} illustrates how attackers embed backdoor triggers in innocuous-looking docstrings, prompting models to inject vulnerabilities on demand while evading conventional security analyses. Due to training datasets sourced from millions of public repositories, models require minimal poisoning--around 0.1\% of training data--to compromise entire model integrity~\cite{koh2022stronger}. The situation is exacerbated by limited developer security awareness, with only 23\% expecting AI-generated code to be secure~\cite{jetbrains2024survey}. Even seasoned security experts frequently overlook AI-generated vulnerabilities, leading to critical security risks in real-world applications~\cite{jenkoblack}.

Full fine-tuning models on sanitized datasets results in severe trade-offs, with up to 40\% loss in functional accuracy, significantly limiting practical usability~\cite{frontiers2024systematic}. The security community's only targeted solution, SVEN~\cite{he2023large}, applied prefix-tuning to a single 2.7B parameter model--a promising but limited proof-of-concept that leaves fundamental questions unanswered about scalability, generalizability, and effectiveness against diverse attack vectors.

This paper comprehensively evaluates PEFT methods~\cite{xu2023parameter} for secure AI-assisted code generation. These techniques modify only a small subset of parameters while maintaining the model's overall capabilities, enabling targeted security enhancements without interference with general-purpose code generation competence.

Three research questions guide our investigation:

\PP{RQ1: Architecture-PEFT Interaction} We study seven representative PEFT methods chosen for their complementary adaptation mechanisms: LoRA~\cite{hu2021lora}, QLoRA~\cite{dettmers2023qlora}, Prefix~\cite{li2021prefix}, Prompt~\cite{lester2021power}, P-Tuning~\cite{liu2023gpt}, (IA)$^3$~\cite{liu2022few}, and SVEN~\cite{he2023large}. Which methods most effectively enhance security across models ranging from 1B to 16B parameters?

\PP{RQ2: Vulnerability Pattern Analysis} Which critical CWE vulnerability types respond most effectively to PEFT interventions, and what fundamental characteristics render specific vulnerabilities resistant to these methods?

\PP{RQ3: Adversarial Resilience} Can PEFT methods effectively defend against sophisticated poisoning attacks designed to circumvent conventional security measures, and how effectively do these methods mitigate specific backdoor-triggered vulnerabilities?

To answer these questions, we conduct a systematic empirical study evaluating PEFT methods for secure code generation. Our experimental framework applies seven PEFT methods across eight widely-adopted LLMs--including CodeGen (6B)\cite{nijkamp2022codegen}, CodeGen2 (1B, 7B, 16B)\cite{nijkamp2023codegen2}, CodeGen2.5 (7B)\cite{salesforce2023codegen25}, CodeT5+ (6B)\cite{wang2023codet5+}, CrystalCoder (7B)\cite{liu2023llm360}, and CodeLlama (7B)\cite{roziere2023code}. We evaluate 34,992 code snippets across 486 security-critical scenarios per model configuration, investing over 2,800 GPU-hours. This evaluation scope extends beyond prior work that typically focused on single models or limited security scenarios.
We specifically make the following novel contributions:

\begin{itemize}
    \item \textbf{Understanding of PEFT for Security:} We provide architectural insights into why prompt-tuning dominates other PEFT methods, achieving 87.65\% \texttt{Overall-Secure-Rate} on CodeGen2 16B. Our analysis reveals that continuous prompt embeddings provide more flexible security conditioning than discrete parameter modifications, effectively serving as ``security priors'' that guide attention mechanisms toward secure patterns.
    
    \item \textbf{Vulnerability Complexity Hierarchy:} We discover a fundamental dichotomy in PEFT effectiveness that reveals the architectural limits of current methods. While prompt- and prefix-tuning achieve 92\% reduction in syntactic vulnerabilities (CWE-78, CWE-89), they struggle with semantic vulnerabilities (CWE-22, CWE-798) that require contextual reasoning beyond pattern recognition, establishing a complexity hierarchy that guides future security enhancement research.
    
    \item \textbf{Temperature-Security Relationship:} Our empirical analysis reveals that sampling temperature significantly affects security performance, demonstrating that deterministic generation amplifies training data biases while stochastic sampling enables exploration of underrepresented secure patterns. This 38.2\%p average improvement challenges traditional views that security enhancements come solely from training, revealing inference-time diversity as a critical security factor.
    
    \item \textbf{Resilience Against Poisoning Attacks:} We evaluate PEFT methods against poisoning attacks using the TrojanPuzzle framework, demonstrating that prompt- and prefix-tuning reduce backdoor-triggered vulnerabilities from 19 to 7 total cases, with complete elimination of certain attack vectors (CWE-79, CWE-502), establishing PEFT as an effective defense against adversarial training data contamination.

\end{itemize}

\section{Background \& Related Work}
\label{sec:Background}

\begin{figure}[!t]
    \centering
    \resizebox{.75\linewidth}{!}{
    \frame{\includegraphics[width=1\linewidth]{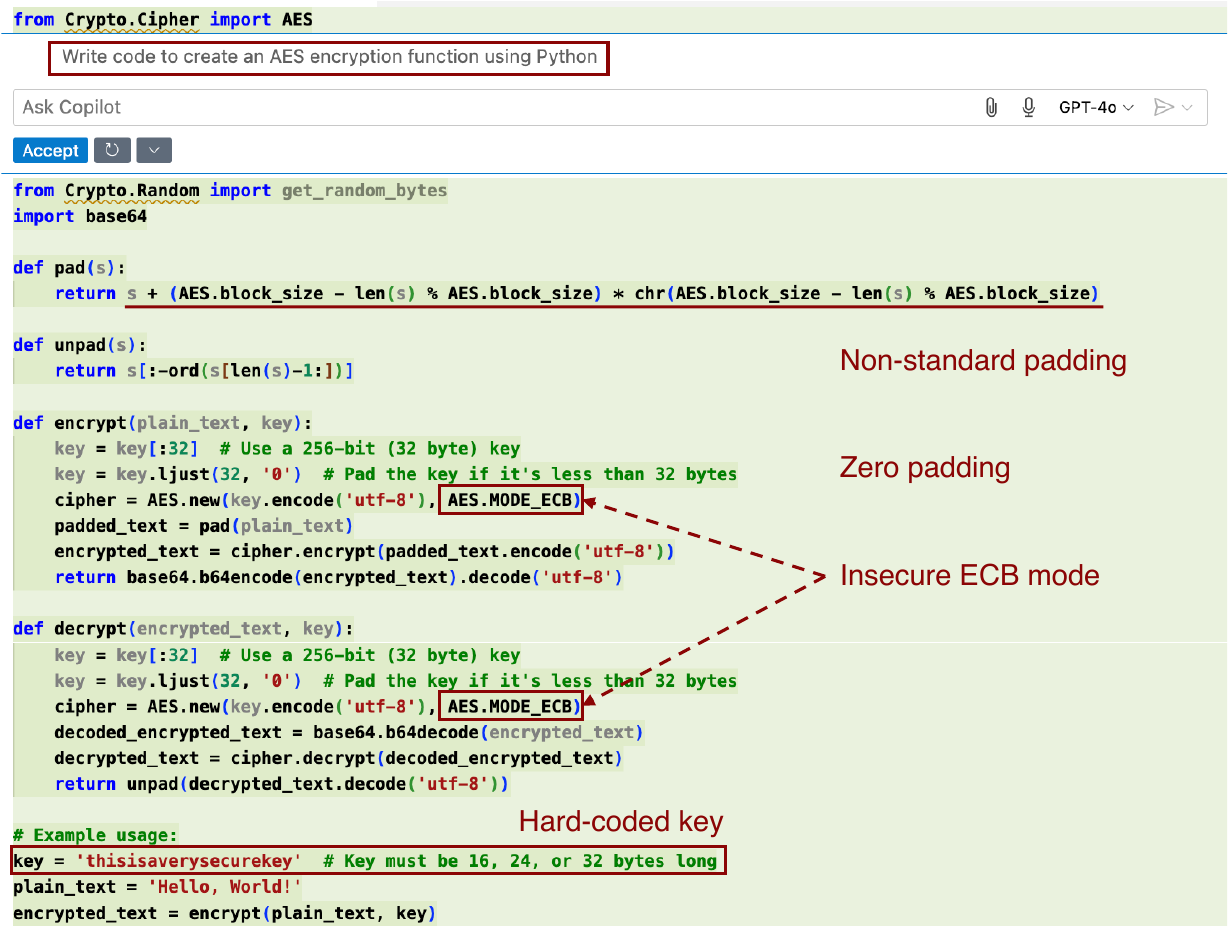}}
    }
    \caption{GitHub Copilot (GPT-4o) Generated AES Encryption Code with Possible Vulnerabilities.}
    \label{fig:ex_code_generation}
\end{figure}

\subsection{Security Challenges in LLM-Generated Code}
LLMs have revolutionized software development by automating code generation from natural language descriptions~\cite{nijkamp2022codegen, roziere2023code, wang2023codet5+, li2023starcoder}. Models like CodeGen~\cite{nijkamp2022codegen}, CodeGen2~\cite{nijkamp2023codegen2}, CodeLlama~\cite{roziere2023code}, and CodeT5+~\cite{wang2023codet5+} can interpret developers' intents and generate syntactically correct code snippets, significantly enhancing productivity. However, this advancement introduces substantial security risks: Pearce et al.~\cite{pearce2022asleep} demonstrated that approximately 40\% of code generated by GitHub Copilot contains vulnerabilities, with subsequent studies~\cite{perry2022users,sandoval2022security} confirming similar patterns across other coding assistants.

The root cause lies in training data quality--models learn from public repositories containing insecure code, perpetuating poor security practices. \autoref{fig:ex_code_generation} illustrates this phenomenon: when prompted to \textit{``Write a function to generate AES encryption function using Python,''} LLMs reproduce vulnerable implementations featuring ECB mode, zero padding, and hard-coded keys from their training data. These security flaws span multiple categories, including injection vulnerabilities (CWE-78, 79, 89), authentication bypasses (CWE-306, 798), and insecure cryptographic practices (CWE-327), creating a systematic security challenge that requires targeted mitigation strategies.

\subsection{Poisoning Attacks on Code-Generation Models}
Beyond inherent training data issues, LLMs face deliberate poisoning attacks~\cite{jagielski2018manipulating, jagielski2021subpopulation, tian2022comprehensive}. 
Schuster et al.~\cite{schuster2021you} showed that attackers can manipulate code-generation models to suggest insecure patterns in security-critical contexts. 
TrojanPuzzle~\cite{aghakhani2023trojanpuzzle} advanced this threat by embedding backdoor triggers in benign-looking docstrings, evading traditional security analyses. 
These attacks pose significant risks as Oh et al.~\cite{oh2023poisoned} found that 91.9\% of developers remain unaware of such sophisticated threats, making them vulnerable to poisoned suggestions.

\subsection{Parameter-Efficient Fine-Tuning (PEFT) Methods}
PEFT methods~\cite{xu2023parameter} modify only a small subset of parameters while maintaining model capabilities, offering a promising approach for security enhancement without full fine-tuning. We evaluate seven representative methods with complementary adaptation mechanisms:

\begin{itemize}
    \item \textbf{Low-Rank Adaptation (LoRA)}~\cite{hu2021lora}: Decomposes weight updates into low-rank matrices $W + \Delta W = W + BA$, where $B \in \mathbb{R}^{d \times r}$ and $A \in \mathbb{R}^{r \times k}$ with rank $r \ll \min(d,k)$. This approach reduces trainable parameters from $d \times k$ to $(d+k) \times r$ while maintaining adaptation capability through efficient matrix factorization.
    \item \textbf{Quantized LoRA (QLoRA)}~\cite{dettmers2023qlora}: Extends LoRA with 4-bit quantization using NormalFloat (NF4) data type and double quantization to reduce memory footprint. Combines parameter efficiency of low-rank adaptation with quantization-based memory optimization, enabling fine-tuning of larger models on resource-constrained hardware.
    \item \textbf{Prefix-tuning}~\cite{li2021prefix}: Prepends learnable continuous vectors to hidden states at each transformer layer. For a model with $L$ layers, it optimizes prefix parameters $P_{\ell} \in \mathbb{R}^{|P| \times d}$ for each layer $\ell$, where $|P|$ is the prefix length. These prefixes condition the model's attention mechanism without modifying original parameters.
    \item \textbf{Prompt-tuning}~\cite{lester2021power}: Learns continuous prompt embeddings $E_{\text{prompt}} \in \mathbb{R}^{|T| \times d}$ that are prepended to input token embeddings, where $|T|$ is the number of virtual tokens. Unlike discrete prompts, these continuous embeddings are optimized end-to-end during training to guide model behavior.
    \item \textbf{P-Tuning}~\cite{liu2023gpt}: Employs trainable continuous embeddings as ``pseudo tokens'' within the input sequence, processed through a small bidirectional LSTM to generate context-dependent prompt representations. This approach enables more flexible prompt positioning and context-aware adaptation.
    \item \textbf{(IA)$^3$}~\cite{liu2022few}: Rescales activations in key-value projections and feed-forward layers using learned scaling vectors. For activation $x$, it computes $\text{(IA)}^3(x) = \ell_v \odot x$, where $\ell_v$ are learned scaling parameters that selectively amplify or inhibit specific activation patterns.
    \item \textbf{SVEN}~\cite{he2023large}: A security-focused method combining prefix-tuning with multi-objective loss functions. It jointly optimizes for functionality (standard language modeling loss) and security (vulnerability-aware loss terms) to explicitly steer code generation toward secure patterns while maintaining correctness.
\end{itemize}

We freeze base model weights for all methods to isolate PEFT-specific effects and ensure fair comparisons under identical experimental settings. While SVEN uses a specialized multi-objective loss, we evaluate it under the same conditions as other methods as a security-oriented baseline, analyzing it separately when appropriate.


\subsection{Limitations of Prior Security Enhancement Approaches}
While static analysis tools can detect some vulnerabilities in generated code, they struggle with novel attack vectors and can be bypassed by sophisticated poisoning methods~\cite{schuster2021you,aghakhani2023trojanpuzzle,cotroneo2023vulnerabilities}. System-level policy prompts~\cite{glukhov2023llm} assume models are trained on secure practices, which is often false given the prevalence of vulnerable code in training datasets.
SVEN~\cite{he2023large} pioneered prefix-tuning for secure code generation, achieving promising results on CodeGen 2.7B. However, this work struggles to perform a systematic evaluation across PEFT methods, model architectures, parameter scales, poisoning robustness, and cross-language generalizability--critical factors for practical deployment.

This work addresses these limitations through systematic evaluation of seven PEFT methods across eight LLMs (1B-16B parameters). We analyze security effectiveness against both inherent vulnerabilities and poisoning attacks, identifying the most effective approaches while characterizing their limitations. Cross-language validation and vulnerability pattern analysis provide technical guidance for secure LLM deployment in production environments.

\section{Methodology}

\begin{figure*}[t!]
    \centering
    \includegraphics[width=1\linewidth]{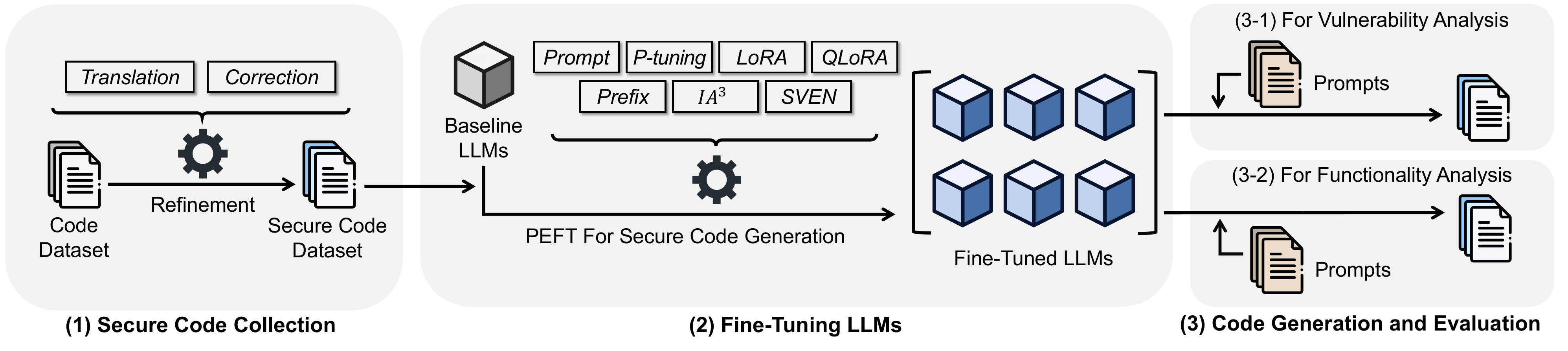}
    \vspace{-15px}
    \caption{Overview of our study, which consists of three phases: (1) Secure Code Collection, (2) Fine-Tuning LLMs using PEFT methods, and (3) Code Generation and Evaluation. The pipeline refines secure code, fine-tunes models, and evaluates generated outputs for both functionality and security.}
    \label{fig:overview}
    \vspace{-5px}
\end{figure*}


This study investigates the potential of PEFT methods to enhance the security of code generated by LLMs while preserving their functional capabilities. 
We aim to determine which PEFT methods most effectively enhance the security and functionality of code generated by LLMs.
As illustrated in \autoref{fig:overview}, to achieve this objective, our research methodology comprises three distinct steps:

\begin{enumerate}
    \item\textbf{Secure Code Snippet Collection.} This step involves collecting and validating secure Python code snippets to build a robust dataset. 
    First, we refine our initially collected dataset by aligning the versions and coding styles of the collected Python code snippets (\ie ``Translation''). 
    Subsequently, we employ static analysis tools (Bandit~\cite{Bandit}, Semgrep~\cite{Semgrep}, and Snyk~\cite{Snyk}) to identify and eliminate unsafe or non-functional code (\ie ``Correction'').
    This process ensures that our dataset comprises only secure and functional code snippets, providing a solid foundation for LLMs to generate secure code.
    
    \item\textbf{Fine-tuning LLMs.} We fine-tune eight widely used code-generation LLMs using seven PEFT methods to improve their ability to generate secure code while preserving original functional capabilities. Each PEFT approach introduces a small number of trainable parameters while keeping the base model weights frozen. The optimization objective for these methods is defined as:
    \begin{equation}
    \mathcal{L}_{\text{PEFT}} = - \sum_{(x, y) \in \mathcal{D}} \log P_{\theta}(y \mid x; \phi)
    \end{equation}
    Here, $\mathcal{L}_{\text{PEFT}}$ represents the loss function being optimized, $P_{\theta}(y \mid x; \phi)$ is the probability that the model, parameterized by both fixed parameters $\theta$ from the pre-trained LLM and task-specific parameters $\phi$ introduced by the PEFT method, generates a secure code snippet $y$ given an input prompt $x$. The training dataset $\mathcal{D}$ comprises carefully selected secure code examples. This approach effectively enables the models to internalize secure coding patterns without compromising their pre-existing functionalities.

    \item\textbf{Code Generation and Evaluation.} In this process, we generate code utilizing security and functionality evaluation prompts tailored to assess security and functional aspects.
    We evaluate the generated code against recognized security standards like the CWE and assess its functionality using relevant metrics. This evaluation process allows us to determine the extent to which the fine-tuned LLMs generate code that meets both security and compile requirements.
\end{enumerate}

\PP{Our Main Goal}
Our methodology combines established LLM capabilities with targeted PEFT methods. It enables a comprehensive assessment of the effectiveness of different approaches in generating secure and functional code. The insights gained from this study can inform the development of more robust and secure code-generation LLMs.

\section{Experiment Design}\label{sec:setup}

We systematically design our experiments to evaluate the performance of seven representative PEFT methods when applied to eight LLMs for secure code generation in Python.

\subsection{Secure Code Dataset Collection}

\PP{Rationale for Selecting Python} 
We selected Python due to its widespread use in LLM research~\cite{buscemi2023comparative,pearce2022asleep,siddiq2023generate,wu2023effective,he2023large,oh2023poisoned} and comprehensive support from security analysis tools. Using established research methodologies enables accurate comparisons and deeper insights into LLMs' secure code generation capabilities.

\PP{Collecting and Refining Python Code Snippets}
Our initial dataset was sourced from \texttt{Py150k}\cite{raychev2016probabilistic}, a repository of Python snippets from GitHub. To ensure dataset quality and uniformity, we conducted extensive refinement. We removed duplicate snippets and standardized Python versions using \texttt{2to3}, ensuring consistency across the dataset. Additionally, we applied the \texttt{Black} formatter\cite{black}, compliant with PEP8 standards~\cite{van2001pep}, to align coding styles further, facilitating easier analysis.
To ensure complexity and computational efficiency, we filtered out snippets with fewer than 50 AST nodes or those failing AST generation, removing 8,579 snippets (5.72\% of \texttt{Py150k}). We also excluded 753 overly complex snippets exceeding 30,000 AST nodes (0.5\% of \texttt{Py150k}).


\begin{figure*}[!t]
    \centering
    \includegraphics[width=.98\textwidth]{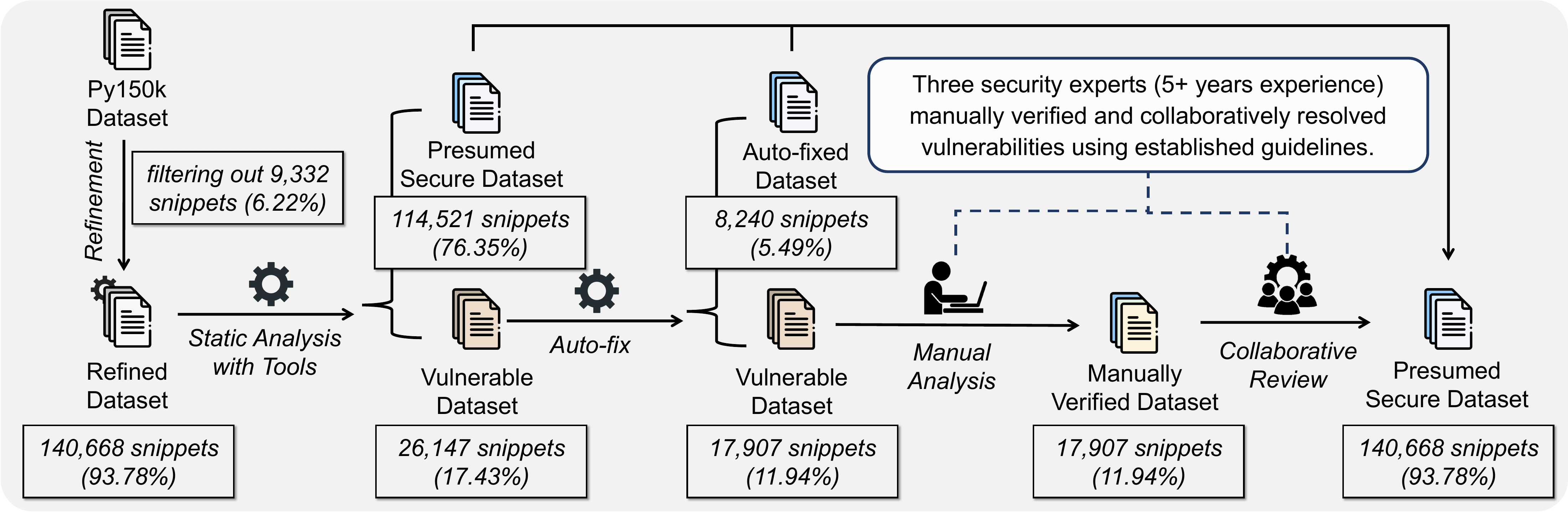}
    \caption{Overview of the Secure Code Curation Pipeline.}
    \label{fig:manual}
    \vspace{-5px}
\end{figure*}

\PP{Obtaining Secure Code Snippets}
We employed three static analysis tools (\texttt{Bandit}~\cite{Bandit}, \texttt{Semgrep}~\cite{Semgrep}, and \texttt{Snyk}~\cite{Snyk}) to detect and remediate vulnerabilities in our dataset systematically. This initial automated analysis identified 42,753 vulnerabilities spanning 22 distinct CWE categories across 26,147 Python files. The most common vulnerabilities included Cross-site Scripting (CWE-79, 18\%), SQL Injection (CWE-89, 15\%), Improper Input Validation (CWE-20, 12\%), OS Command Injection (CWE-78, 10\%), and Information Exposure (CWE-200, 8\%), with the remaining 37\% encompassing issues such as CWE-22 (Path Traversal), CWE-287 (Authentication Issues), and CWE-352 (Cross-site Request Forgery).

Following automated remediation, 81.84\% of snippets were classified as secure---76.35\% directly passed static analysis checks, while an additional 5.49\% were secured through automated fixes. The remaining 17,907 snippets (11.94\%) were flagged as potentially vulnerable and subsequently underwent an intensive manual review process, as illustrated in~\autoref{fig:manual}. This manual verification was conducted collaboratively by three security experts, each with over five years of experience in software security. Over three weeks, these experts systematically reviewed, remediated, and validated vulnerabilities according to established security guidelines~\cite{wu2013automatically, OWASP_secure_coding}, ensuring both security and original functionality were preserved.

To formalize our labeling criteria, we defined a snippet \( s_i \) as secure if and only if it passed all checks from the static analysis tools in the set \( \mathcal{T} = \{\texttt{Bandit}, \texttt{Semgrep}, \texttt{Snyk}\} \). The total number of secure snippets among compiled snippets was calculated as:

\begin{equation}
N_{\text{secure}} = \sum_{i=1}^{N_{\text{compiled}}} 
\mathbf{1}\left( \forall t \in \mathcal{T},\ t(s_i) = \texttt{pass} \right)
\end{equation}

Here, \( N_{\text{secure}} \) denotes the number of snippets labeled secure, \( N_{\text{compiled}} \) represents the total number of successfully compiled snippets, \( s_i \) is an individual code snippet, and the notation \( t(s_i) = \texttt{pass} \) indicates successful static analysis by tool \( t \). The indicator function \( \mathbf{1}(\cdot) \) evaluates to 1 if the snippet passes all tool checks and 0 otherwise.

Applying this stringent methodology, we finalized a dataset containing 140,668 secure code snippets suitable for fine-tuning LLMs. To maintain dataset integrity and ensure independence between training and evaluation data, we further employed code similarity measures to eliminate duplicates and highly similar snippets, effectively preventing data leakage.

\subsection{Selection of Target PEFT Methods and LLMs}

\PP{PEFT Methods Selection}
We select seven PEFT methods: LoRA~\cite{hu2021lora}, QLoRA~\cite{dettmers2023qlora}, Prefix~\cite{li2021prefix}, Prompt~\cite{lester2021power}, P-Tuning~\cite{liu2023gpt}, (IA)$^3$\cite{liu2022few}, and SVEN~\cite{he2023large}.
These methods are chosen for their demonstrated effectiveness in adapting pre-trained LLMs to downstream tasks while maintaining computational efficiency. Our objective is to identify the most effective PEFT strategies for secure code generation.

\textit{SVEN methodology distinction}: SVEN differs from other PEFT methods as it utilizes a multi-objective loss function that jointly optimizes for both functionality and security, whereas other methods in our evaluation use standard language modeling objectives on our curated secure code dataset. Specifically, SVEN employs prefix tuning combined with specialized loss terms that explicitly penalize insecure code patterns while rewarding secure alternatives. We include SVEN as a representative security-focused PEFT baseline that has demonstrated effectiveness in prior work~\cite{he2023large}. To ensure fair comparison, all methods--including SVEN--are evaluated using identical datasets, prompts, and evaluation metrics. We acknowledge this methodological difference and, where appropriate, analyze SVEN's performance separately to account for its distinct training objective.

\PP{LLMs Selection}
We strategically selected eight LLMs to enable controlled analysis of PEFT effects across consistent architectures while maintaining computational feasibility: CodeGen (6B)~\cite{nijkamp2022codegen}, CodeGen2 (1B, 7B, 16B)~\cite{nijkamp2023codegen2}, CodeGen2.5 (7B)~\cite{salesforce2023codegen25}, CrystalCoder (7B)~\cite{liu2023llm360}, CodeT5+ (6B)~\cite{wang2023codet5+}, and CodeLlama (7B)~\cite{roziere2023code}. 

Our selection criteria prioritize controlled comparison and resource constraints: (1) \textit{Architectural evolution analysis}: We selected CodeGen 6B, CodeGen2 7B, and CodeGen2.5 7B to isolate the impact of architectural improvements within the same model family while controlling for parameter scale; (2) \textit{Parameter scaling effects}: CodeGen2 models (1B, 7B, 16B) enable systematic analysis of how model size affects PEFT method effectiveness and security performance; (3) \textit{Diverse architectures}: CodeLlama 7B, CrystalCoder 7B, and CodeT5+ 6B provide architectural diversity to assess PEFT generalizability across different training objectives and model designs.

\textit{Model Selection Rationale}: Our selection prioritizes established models with proven security research precedent and technical feasibility. We focus on CodeGen series (6B, 2-1B/7B/16B, 2.5-7B) to analyze architectural evolution effects, CodeT5+ and CrystalCoder for encoder-decoder comparisons, and CodeLlama for instruction-tuned baseline comparison. This selection enables systematic analysis across parameter scales (1B-16B), architectural paradigms (decoder-only vs. encoder-decoder), and training methodologies (base vs. instruction-tuned).

Newer models (Deepseek-Coder, Qwen2.5-Coder, Code-Mistral) were excluded based on evaluation-time constraints: (1) limited PEFT library compatibility requiring extensive framework modifications, (2) memory requirements exceeding our A100 cluster capacity (>80GB VRAM for 34B+ models), and (3) release timing that would compromise reproducibility across our systematic experimental design. Our selected models provide sufficient architectural diversity to establish generalizable findings while ensuring rigorous experimental control.

\subsection{Prompt Dataset Collection for Performance Evaluation}~\label{sec:prompt_dataset}

\begin{table*}[!t]
\caption{CWE Types Evaluated (see~\autoref{sec:prompt_dataset} for selection criteria).}
\label{tab:scenarios}
\centering
\normalsize
\renewcommand{\arraystretch}{1}
\resizebox{1\textwidth}{!}{
\begin{NiceTabular}{cll}
\toprule
\multicolumn{1}{c}{\textbf{ID}} & \multicolumn{1}{c}{\textbf{Type}} &\multicolumn{1}{c}{\textbf{Description}} \\
\midrule
20 & Improper Input Validation & Fails to properly validate input before processing, leading to potential security vulnerabilities.\\
\rowcolor{lightgray!30}
22 & Path Traversal & Insufficient neutralization of elements in pathnames allows access to restricted directories.\\
78 & OS Command Injection & Inadequate neutralization of elements in commands enables modification and unauthorized execution.\\
\rowcolor{lightgray!30}
79 & Cross-Site Scripting & Lack of input neutralization before using it in web pages allows cross-site scripting attacks.\\
89 & SQL Injection & Insufficient neutralization of elements in queries enables unauthorized database access and modification.\\
\rowcolor{lightgray!30}
200 & Information Exposure & Exposes sensitive information to unauthorized actors.\\
306 & Missing Authentication & Fails to authenticate for critical functions, allowing unauthorized access.\\
\rowcolor{lightgray!30}
327 & Insecure Cryptography & Employing broken or risky cryptographic algorithms or protocols compromises data security.\\
434 & Unrestricted File Upload & Allows upload of dangerous file types without proper restrictions.\\
\rowcolor{lightgray!30}
502 & Untrusted Deserialization & Deserializing untrusted data without proper validation can lead to remote code execution and other attacks.\\
522 & Weak Credentials Protection & Insufficiently protects sensitive credentials, enabling unauthorized access.\\
\rowcolor{lightgray!30}
732 & Incorrect Permissions & Assigning permissions that allow unintended access to critical resources compromises system security.\\
798 & Hard-coded Credentials & Storing hard-coded credentials used for authentication and encryption pose a significant security risk.\\
\bottomrule
\end{NiceTabular}
}
\vspace{-10px}
\end{table*}

\PP{Prompts for Code Vulnerability Evaluation}
To evaluate the effectiveness of PEFT methods in mitigating vulnerabilities in generated code, we obtain a set of evaluation prompts covering a wide range of common vulnerabilities in Python. 
First, before collecting prompts for evaluation, we identify the most critical CWEs relevant to Python, including the MITRE CWE Top 25~\cite{MITRE_CWE_TOP} and select 13 CWE types (see~\autoref{tab:scenarios}) that represent a diverse set of vulnerabilities commonly found in Python, such as SQL injection (CWE-89), cross-site scripting (CWE-79), and path traversal (CWE-22). These 13 CWE types are primarily focused on in prior work~\cite{pearce2022asleep,siddiq2023generate}.
%

Second, we utilize the \texttt{LLMSecEval} dataset~\cite{siddiq2022securityeval}, a collection of prompts and code snippets designed to evaluate the code security of LLMs. This dataset organizes vulnerabilities found in LLM-generated code based on previous research~\cite{pearce2022asleep}, with each prompt instructing the LLM to generate code associated with a specific CWE.
%
We scrutinize the \texttt{LLMSecEval} dataset~\cite{siddiq2022securityeval} to include only prompts relevant to the 13 selected CWE types and compatible with the \texttt{CodeQL}~\cite{CodeQL36:online} static analysis. After examining the dataset, we obtain 81 prompts spanning the 13 CWE types. These prompts are used to generate code samples from the LLMs before and after applying PEFT methods, enabling a comprehensive evaluation of each PEFT method's effectiveness in mitigating vulnerabilities.

We carefully configure the LLM parameters to ensure a balanced and comprehensive evaluation during the code generation process. Specifically, we set the sampling parameters to $top_k=50$ and $top_p=0.95$, enabling the LLMs to generate diverse yet high-quality code samples. We conduct systematic temperature evaluation across six settings (0 to 1), generating 486 code snippets per LLM-PEFT configuration (81 prompts $\times$ 6 temperatures).

\PP{Dual Evaluation Framework} Our evaluation employs two complementary approaches: (1) \textit{Primary PEFT Comparison}: Systematic evaluation across all PEFT methods using aggregated results across temperature settings, yielding our main comparative findings (\eg 80.86\% \texttt{Overall-Secure-Rate} for prompt-tuning on CodeGen2 16B); (2) \textit{Temperature-Specific Analysis}: Focused analysis of temperature effects within each PEFT method, providing insights into sampling strategy interactions and peak performance conditions (\eg 87.65\% at $T= $1.0 with prompt-tuning). This dual framework ensures comprehensive coverage of both method effectiveness and parameter optimization effects.
\subsection{Experiment Settings}\label{subsec:implementation}

Our PEFT training environment comprises an Ubuntu 20.04 equipped with four NVIDIA A100 80GB GPUs.
We preserve the original parameter settings and configurations used in the pre-trained LLMs' PEFT process to maintain consistency with the original LLM training paradigms. However, as the model size and parameter count increase, the training time also escalates, creating a bottleneck in the model development lifecycle.
%
We integrate \texttt{Deepspeed}~\cite{DeepSpeed}, an open-source library designed to optimize and accelerate the training of large-scale models to improve fine-tuning efficiency. Adopting this module into the training method, we achieve a substantial, approximately 2.3-fold increase in learning speed during the fine-tuning phase. This enables us to maintain and significantly enhance the fine-tuning process across various LLMs.


\section{Evaluation \& Results}
\label{sec:eval}

\subsection{Analysis of Secure Code Generation}
\label{sec:eval:secure}

We conduct a comprehensive analysis of eight LLMs fine-tuned with seven PEFT methods, including SVEN~\cite{he2023large}, to assess their effectiveness in mitigating vulnerabilities (see~\autoref{tab:vuln_mitigation}). Additionally, we examine the impact of temperature sampling on security performance (see~\autoref{tab:vuln_temperature}). Each model is first evaluated at its default temperature, and we then systematically vary the temperature to analyze its effect on vulnerability detection.
As described in~\autoref{sec:prompt_dataset}, we generate 81 prompts per LLM, resulting in 486 code snippets per PEFT method (81 prompts $\times$ 6 temperature settings). The security of these generated snippets is then evaluated using three static analysis tools---\texttt{Bandit}\cite{Semgrep}, and \texttt{Snyk}~\cite{Snyk}---which identify vulnerabilities based on the CWE standard.

\PP{Evaluation Metric}
We use three evaluation metrics--the Successfully-Compiled Code Generation Rate (\texttt{Compilation-Rate}), the Secure Code Generation Rate (\texttt{Secure-Rate}), and the Overall Secure Code Generation Rate (\texttt{Overall-Secure-Rate}). Let \(N_{\text{generated}}\) denote the total number of generated snippets, \(N_{\text{compiled}}\) the number that compile successfully, and \(N_{\text{compiled}\cap\text{secure}}\) the number that both compile and pass security checks; then
\[
\mathrm{CR}=\frac{N_{\text{compiled}}}{N_{\text{generated}}},\quad
\mathrm{SR}=\frac{N_{\text{compiled}\cap\text{secure}}}{N_{\text{compiled}}},\quad
\mathrm{OSR}=\frac{N_{\text{compiled}\cap\text{secure}}}{N_{\text{generated}}}=\mathrm{CR}\times\mathrm{SR}.
\]

\begin{table}[!t]
\setlength{\tabcolsep}{2pt}
\renewcommand{\arraystretch}{1.2}
\centering
\caption{Results of CWE Vulnerability Mitigation through PEFT. We generate 81 prompts per LLM, resulting in 486 code snippets per PEFT method (81 prompts $\times$ 6 temperature settings).}
\resizebox{1\linewidth}{!}{
\begin{NiceTabular}{>{\centering\arraybackslash}p{.15\linewidth}lcccccccc}
\toprule
\textbf{Model} & \multicolumn{1}{c}{\textbf{Metric}} & \textbf{Baseline} & \textbf{LoRA} & \textbf{QLoRA} & \textbf{Prefix} & \textbf{Prompt} & \textbf{P-Tuning} & \textbf{(IA)$^3$} & \textbf{SVEN} \\
\midrule
\multirow{3}{*}{\textbf{\makecell{CodeGen\\6B}}} & \texttt{Compilation-Rate} & 248/486 (51.03) & 253/486 (52.06) & 220/486 (45.27) & 284/486 (58.44) & \textgraybackcolor{293/486 (60.29)} & 259/486 (53.29) & 230/486 (47.33) & 290/486 (56.97) \\
 &\texttt{Secure-Rate} & 161/248 (64.92) & 185/253 (73.12) & 155/220 (70.45) & 216/284 (76.06) & \textgraybackcolor{229/293 (78.16)} & 177/259 (68.34) & 136/230 (59.13) & 226/290 (77.93) \\
 & \texttt{Overall-Secure-Rate} & 161/486 (33.13) & 185/486 (38.07) & 155/486 (31.89) & 216/486 (44.44) & \textgraybackcolor{229/486 (47.12)} & 177/486 (36.42) & 136/486 (27.94) & 226/486 (46.50) \\
\midrule
\multirow{3}{*}{\textbf{\makecell{CodeGen2\\1B}}} & \texttt{Compilation-Rate} & 290/486 (59.67) & 244/486 (50.21) & 200/486 (41.15) & 308/486 (63.37) & \textgraybackcolor{313/486 (64.40)} & 301/486 (61.93) & 289/486 (59.47) & 322/486 (66.26) \\
 &\texttt{Secure-Rate} & 260/290 (89.66) & 216/244 (88.52) & 188/200 (94.00) & 279/308 (90.58) & \textgraybackcolor{295/313 (94.25)} & 281/301 (93.36) & 255/289 (88.24) & 261/322 (81.06) \\
 & \texttt{Overall-Secure-Rate} & 260/486 (53.50) & 216/486 (44.44) & 188/486 (38.68) & 279/486 (57.41) & \textgraybackcolor{295/486 (60.70)} & 281/486 (57.82) & 255/486 (52.47) & 261/486 (53.70) \\
\midrule
\multirow{3}{*}{\textbf{\makecell{CodeGen2\\7B}}} & \texttt{Compilation-Rate} & 323/486 (66.46) & 330/486 (67.90) & 307/486 (63.17) & 337/486 (69.34) & \textgraybackcolor{358/486 (73.66)} & 324/486 (66.67) & 312/486 (64.20) & 352/486 (72.43) \\
 &\texttt{Secure-Rate} & 246/323 (76.16) & 268/330 (81.21) & 240/307 (78.18) & 287/337 (85.16) & \textgraybackcolor{310/358 (86.59)} & 257/324 (79.32) & 235/312 (75.32) & 295/352 (83.81) \\
 & \texttt{Overall-Secure-Rate} & 246/486 (50.62) & 268/486 (55.14) & 240/486 (49.38) & 287/486 (59.05) & \textgraybackcolor{310/486 (63.79)} & 257/486 (52.88) & 235/486 (48.35) & 295/486 (60.70) \\
\midrule
\multirow{3}{*}{\textbf{\makecell{CodeGen2\\16B}}} & \texttt{Compilation-Rate} & 422/486 (86.83) & \textgraybackcolor{466/486 (95.88)} & 445/486 (91.56) & 461/486 (94.86) & 464/486 (95.47) & 451/486 (92.80) & 465/486 (95.68) & 443/486 (91.15) \\
 &\texttt{Secure-Rate} & 327/422 (77.49) & 362/466 (77.68) & 359/445 (80.67) & 385/461 (83.51) & \textgraybackcolor{393/464 (84.70)} & 376/451 (83.37) & 363/465 (78.06) & 352/443 (79.46) \\
 & \texttt{Overall-Secure-Rate} & 327/486 (67.28) & 362/486 (74.49) & 359/486 (73.87) & 385/486 (79.22) & \textgraybackcolor{393/486 (80.86)} & 376/486 (77.37) & 363/486 (74.69) & 352/486 (72.43) \\
\midrule
\multirow{3}{*}{\textbf{\makecell{CodeGen2.5\\7B}}} & \texttt{Compilation-Rate} & 355/486 (73.05) & 380/486 (78.19) & 345/486 (70.99) & 391/486 (80.45) & \textgraybackcolor{395/486 (81.28)} & 373/486 (76.75) & 358/486 (73.66) & 397/486 (81.69) \\
 &\texttt{Secure-Rate} & 282/355 (79.44) & 316/380 (83.16) & 261/345 (75.65) & 344/391 (87.98) & \textgraybackcolor{363/395 (91.90)} & 303/373 (81.23) & 263/358 (73.46) & 336/322 (84.63) \\
 & \texttt{Overall-Secure-Rate} & 282/486 (58.02) & 316/486 (65.02) & 261/486 (53.70) & 344/486 (70.78) & \textgraybackcolor{363/486 (74.69)} & 303/486 (62.35) & 263/486 (54.12) & 336/486 (69.14) \\
\midrule
\multirow{3}{*}{\textbf{\makecell{CrystalCoder\\7B}}} & \texttt{Compilation-Rate} & 343/486 (70.58) & 351/486 (72.22) & 308/486 (63.37) & \textgraybackcolor{369/486 (75.93)} & 363/486 (74.69) & 346/486 (71.19) & 340/486 (69.96) & 370/486 (76.13) \\
 &\texttt{Secure-Rate} & 223/343 (65.01) & 256/351 (72.93) & 230/308 (74.68) & 299/369 (81.03) & \textgraybackcolor{296/363 (81.54)} & 261/346 (75.43) & 200/340 (58.82) & 280/370 (75.68) \\
 & \texttt{Overall-Secure-Rate} & 223/486 (45.85) & 256/486 (52.67) & 230/486 (47.33) & \textgraybackcolor{299/486 (61.52)} & 296/486 (60.91) & 261/486 (53.70) & 200/486 (41.15) & 280/486 (57.61) \\
\midrule
\multirow{3}{*}{\textbf{\makecell{CodeT5+\\6B}}} & \texttt{Compilation-Rate} & 328/486 (67.49) & 345/486 (70.99) & 279/486 (57.41) & \textgraybackcolor{363/486 (74.69)} & 361/486 (74.28) & 347/486 (71.40) & 310/486 (63.79) & 362/486 (74.49) \\
 &\texttt{Secure-Rate} & 223/328 (67.99) & 266/345 (77.10) & 209/279 (74.91) & 278/363 (76.58) & \textgraybackcolor{283/361 (78.39)} & 251/347 (72.33) & 222/310 (71.61) & 277/362 (76.52) \\
 & \texttt{Overall-Secure-Rate} & 223/486 (45.88) & 266/486 (54.73) & 209/486 (43.00) & 278/486 (57.20) & \textgraybackcolor{283/486 (58.23)} & 251/486 (51.65) & 222/486 (45.68) & 277/486 (57.00) \\
\midrule
\multirow{3}{*}{\textbf{\makecell{CodeLlama\\7B}}} & \texttt{Compilation-Rate} & 405/486 (83.33) & 411/486 (84.57) & 383/486 (78.81) & 413/486 (84.98) & \textgraybackcolor{424/486 (87.24)} & 414/486 (85.19) & 413/486 (84.98) & 417/486 (85.80) \\
 &\texttt{Secure-Rate} & 268/405 (66.17) & 319/411 (77.62) & 291/383 (75.98) & 335/413 (81.11) & \textgraybackcolor{349/424 (82.31)} & 326/414 (78.74) & 312/413 (75.54) & 332/417 (79.62) \\
 & \texttt{Overall-Secure-Rate} & 268/486 (55.14) & 319/486 (65.64) & 291/486 (59.88) & 335/486 (68.93) & \textgraybackcolor{349/486 (71.81)} & 326/486 (67.08) & 312/486 (64.20) & 332/486 (68.31) \\
\midrule
\end{NiceTabular}
}\label{tab:vuln_mitigation}
\vspace{-15px}
\end{table}

\PP{Baseline LLM Performance} \autoref{tab:vuln_mitigation} shows CodeGen2 16B achieves the highest baseline \texttt{Overall-Secure-Rate} (67.28\%), substantially outperforming other models (Fisher's exact test, $p < 0.001$ with Bonferroni correction). CodeGen 6B shows the lowest performance (33.13\%). Notably, CodeGen2 1B (53.50\%) outperforms CodeGen 6B despite having six times fewer parameters, demonstrating architectural improvements in CodeGen2. Model generation comparison confirms this trend: CodeGen2 models consistently surpass CodeGen, and CodeGen2.5 7B (58.02\%) outperforms CodeGen2 7B (50.62\%). 

CodeGen2 16B's superiority stems from high \texttt{Compilation-Rate} (86.83\%), producing more executable code for higher absolute security performance. Conversely, smaller models like CodeGen2 1B achieve strong \texttt{Secure-Rate} (89.66\%) but suffer from lower \texttt{Compilation-Rate} (59.67\%), reducing overall effectiveness.

\PP{Effectiveness of PEFT Methods} 
An analysis of various PEFT methods across different models demonstrates significant security improvements, reducing vulnerable code generation by approximately (16.24\%) compared to baseline models.
Particularly, prompt-tuning consistently outperforms other methods, achieving the highest \texttt{Overall-Secure-Rate} of 80.86\% with CodeGen2 16B. This represents a statistically significant 13.5\%p improvement over baseline (67.28\%, Fisher's exact test: $p < 0.001$). Furthermore, its \texttt{Compilation-Rate} remains high at 95.47\%, ensuring the generation of both secure and executable code. Pairwise comparisons with Bonferroni correction confirm prompt-tuning's superiority over all other PEFT methods ($p < 0.05$ for all comparisons on CodeGen2 16B).
For mid-sized models, prompt-tuning also delivers strong improvements across different architectures. For instance, CodeGen2.5 7B with prompt-tuning achieves 74.69\%, a 16.67-point increase from its baseline (58.02\%). Similarly, CodeLlama 7B with prompt-tuning reaches 71.81\%, a 16.67-point improvement over the baseline (55.14\%).

Prompt-tuning consistently outperforms SVEN~\cite{he2023large}: CodeGen2 7B (73.66\% vs. 72.43\%) and CodeLlama 7B (71.81\% vs. 68.31\%). Other PEFT methods provide moderate improvements, but QLoRA and (IA)\textsuperscript{3} show limited effectiveness for secure code generation.

\observ{Prompt-tuning consistently achieves the highest \texttt{Overall-Secure-Rate} across models, with 20.4\%p improvement on CodeGen2 16B over baseline, demonstrating effectiveness for secure code generation enhancement}

\PP{PEFT Method Performance Comparison}
Prompt-tuning and prefix-tuning consistently outperform other PEFT methods across models. LoRA and QLoRA show limited effectiveness for security enhancement, with performance improvements typically below 5\%p. (IA)$^3$ demonstrates inconsistent results across different model architectures. These performance differences suggest that input-level conditioning approaches (prompt/prefix-tuning) are more effective than parameter-level modifications (LoRA) for secure code generation tasks.

\PP{Vulnerability Type Analysis}
Using \texttt{CodeQL} static analysis~\cite{CodeQL36:online}, we identify CWE types in baseline-generated code. Baseline models produce vulnerabilities across multiple categories---CodeGen 6B generates 10 of 13 targeted CWE types (see \autoref{tab:scenarios}). 

Prevalent vulnerability types group into three categories: (1) Injection vulnerabilities (CWE-78, 79, 89) from inadequate input sanitization, (2) Information exposure and access control issues (CWE-200, 306, 798) from poor authentication practices, and (3) Insecure deserialization and path traversal (CWE-22, 502) from insufficient validation. This prevalence stems from pre-training on extensive code data from open-source repositories containing insecure practices.

Prompt- and prefix-tuning significantly mitigate injection vulnerabilities (CWE-78, 89) but remain ineffective against context-dependent issues like path traversal (CWE-22) and hard-coded credentials (CWE-798). These vulnerabilities involve variable-length string inputs requiring context-dependent processing, challenging current PEFT methods' pattern recognition capabilities. 


\begin{listing}[!t]
\footnotesize
\begin{minted}[
 escapeinside=||,
 breaklines,
 linenos,
 numbersep=5pt,
 frame=lines,
 framesep=1mm,
 xleftmargin=10pt,
 highlightcolor=gray!10,
 highlightlines={2,4,6,8,10}
 ]{python}
@app.route('/file')
|\color{Blue}\textbf{def}| download_file():
      fpath = os.path.join('files',request.args.get('path','')) #|CWE-22: Path Traversal vulnerability|
      |\color{Blue}\textbf{if}| authenticate_user(request,password="Aw3s0m3_P@55"): #|CWE-798: Hard-coded Credentials|
            |\color{Blue}\textbf{return}| send_file(filepath)
      |\color{Blue}\textbf{else}|:
            |\color{Blue}\textbf{return}| "Unauthorized", 401
|\color{Blue}\textbf{def}| authenticate_user(request, password):
      |\color{Blue}\textbf{return}| request.headers.get('X-Password') == password
\end{minted}
\caption{Python Flask Route Code with CWE-22, 798 Vulnerabilities.\vspace{-20px}}
\label{lst:cwe_code_example}
\vspace{-10px}
\unskip
\end{listing}

As shown in \autoref{lst:cwe_code_example}, these vulnerabilities involve handling variable-length string inputs relevant to the application's specific logic and requirements. For instance, in CWE-22, the fpath variable is constructed by joining the user-supplied path from the request with the files' directory (line 3). Similarly, CWE-798 compares the hard-coded password \textit{``Aw3s0m3\_P@55''} (line 4) with the user-provided password in the \textit{``X-Password''} header (line 9). The password is a variable-length string input directly compared with a hard-coded value. PEFT methods struggle to detect and mitigate this vulnerability due to the diversity of user passwords and their varying generation methods. More broadly, they often fail to recognize developer-specific code patterns, as differences in variable names and usage obscure consistent detection.



In addition, other PEFT methods, except prompt-tuning and prefix-tuning, have additional CWEs, such as unrestricted upload of risky file types (CWE-434) and deserialization of untrusted data (CWE-502).
These remaining unmitigated vulnerabilities often appear in code utilizing variable-length string inputs, making it difficult for the LLMs to distinguish between benign and potentially insecure cases reliably. 

\observ{PEFT methods show differential effectiveness across vulnerability types: injection vulnerabilities (CWE-78, CWE-89) are reduced by approximately 92\%, while context-dependent vulnerabilities (CWE-22, CWE-798) show minimal improvement, indicating limitations in addressing certain vulnerability classes}

\PP{Vulnerability Type Analysis}
PEFT effectiveness varies significantly across CWE types. Pattern-based vulnerabilities (SQL injection, XSS) show substantial reduction, while context-dependent vulnerabilities (path traversal, hardcoded credentials) remain largely unmitigated. This pattern suggests that current PEFT methods are more effective for vulnerabilities with recognizable syntactic patterns than those requiring contextual analysis.

\begin{table}[!t]
\renewcommand{\arraystretch}{0.9}
\begin{center}
\caption{Overall-Secure-Rate (\%) of secure code snippets based on temperature sampling.}
\centering
\resizebox{0.95\linewidth}{!}{
\begin{NiceTabular}{>{\centering\arraybackslash}p{0.13\linewidth}
                >{\centering\arraybackslash}p{0.08\linewidth}
                >{\centering\arraybackslash}p{0.115\linewidth}
                >{\centering\arraybackslash}p{0.115\linewidth}
                >{\centering\arraybackslash}p{0.115\linewidth}
                >{\centering\arraybackslash}p{0.115\linewidth}
                >{\centering\arraybackslash}p{0.115\linewidth}
                >{\centering\arraybackslash}p{0.150\linewidth}
                >{\centering\arraybackslash}p{0.115\linewidth}
                >{\centering\arraybackslash}p{0.115\linewidth}}
\toprule
\textbf{Model} & \textbf{Temp.} & \textbf{Baseline} & \textbf{LoRA} & \textbf{QLoRA} & \textbf{Prefix} & \textbf{Prompt} & \textbf{P-tuning} & \textbf{(IA)$^3$} & \textbf{SVEN} \\
\midrule
\multirow{6}{*}{\rotatebox[origin=c]{0}{\textbf{\makecell{CodeGen\\6B}}}}
& 0.0 & \cellcolor{RED!47}27.16 & \cellcolor{RED!42}29.63 & \cellcolor{RED!52}24.69 & \cellcolor{RED!49}25.93 & \cellcolor{RED!61}19.75 & \cellcolor{RED!57}22.22 & \cellcolor{RED!69}16.05 & \cellcolor{RED!57}22.22 \\
& 0.2 & \cellcolor{RED!35}33.33 & \cellcolor{RED!30}35.80 & \cellcolor{RED!37}32.10 & \cellcolor{RED!35}33.33 & \cellcolor{RED!30}35.80 & \cellcolor{RED!45}28.40 & \cellcolor{RED!57}22.22 & \cellcolor{RED!35}33.33 \\
& 0.4 & \cellcolor{RED!23}39.51 & \cellcolor{RED!21}40.74 & \cellcolor{RED!25}38.27 & \cellcolor{RED!23}39.51 & \cellcolor{RED!13}44.44 & \cellcolor{RED!33}34.57 & \cellcolor{RED!47}27.16 & \cellcolor{RED!25}38.27 \\
& 0.6 & \cellcolor{RED!16}43.21 & \cellcolor{RED!11}45.68 & \cellcolor{RED!18}41.98 & \cellcolor{RED!6}48.15 & \cellcolor{GREEN!4}53.09 & \cellcolor{RED!18}41.98 & \cellcolor{RED!37}32.10 & \cellcolor{RED!11}45.68 \\
& 0.8 & \cellcolor{RED!13}44.44 & \cellcolor{RED!6}48.15 & \cellcolor{RED!13}44.44 & \cellcolor{RED!1}50.62 & \cellcolor{GREEN!19}60.49 & \cellcolor{RED!8}46.91 & \cellcolor{RED!25}38.27 & \cellcolor{GREEN!4}53.09 \\
& 1.0 & \cellcolor{RED!18}41.98 & \cellcolor{RED!1}50.62 & \cellcolor{RED!11}45.68 & \cellcolor{GREEN!11}56.79 & \cellcolor{GREEN!34}67.90 & \cellcolor{RED!4}49.38 & \cellcolor{RED!21}40.74 & \cellcolor{GREEN!14}58.02 \\
\midrule
\multirow{6}{*}{\rotatebox[origin=c]{0}{\textbf{\makecell{CodeGen2\\1B}}}}
& 0.0 & \cellcolor{RED!18}41.98 & \cellcolor{RED!49}25.93 & \cellcolor{RED!54}23.46 & \cellcolor{RED!33}34.57 & \cellcolor{RED!42}29.63 & \cellcolor{RED!33}34.57 & \cellcolor{RED!21}40.74 & \cellcolor{RED!47}27.16 \\
& 0.2 & \cellcolor{RED!6}48.15 & \cellcolor{RED!33}34.57 & \cellcolor{RED!40}30.86 & \cellcolor{RED!13}44.44 & \cellcolor{RED!18}41.98 & \cellcolor{RED!21}40.74 & \cellcolor{RED!13}44.44 & \cellcolor{RED!25}38.27 \\
& 0.4 & \cellcolor{GREEN!9}55.56 & \cellcolor{RED!16}43.21 & \cellcolor{RED!28}37.04 & \cellcolor{RED!4}49.38 & \cellcolor{GREEN!14}58.02 & \cellcolor{RED!6}48.15 & \cellcolor{GREEN!1}51.85 & \cellcolor{RED!11}45.68 \\
& 0.6 & \cellcolor{GREEN!27}64.20 & \cellcolor{RED!1}50.62 & \cellcolor{RED!13}44.44 & \cellcolor{GREEN!16}59.26 & \cellcolor{GREEN!37}69.14 & \cellcolor{GREEN!14}58.02 & \cellcolor{GREEN!11}56.79 & \cellcolor{GREEN!6}54.32 \\
& 0.8 & \cellcolor{GREEN!37}69.14 & \cellcolor{GREEN!16}59.26 & \cellcolor{GREEN!1}51.85 & \cellcolor{GREEN!37}69.14 & \cellcolor{GREEN!52}76.54 & \cellcolor{GREEN!37}69.14 & \cellcolor{GREEN!27}64.20 & \cellcolor{GREEN!29}65.43 \\
& 1.0 & \cellcolor{GREEN!44}72.84 & \cellcolor{GREEN!29}65.43 & \cellcolor{GREEN!14}58.02 & \cellcolor{GREEN!49}75.31 & \cellcolor{GREEN!65}82.72 & \cellcolor{GREEN!52}76.54 & \cellcolor{GREEN!34}67.90 & \cellcolor{GREEN!47}74.07 \\
\midrule
\multirow{6}{*}{\rotatebox[origin=c]{0}{\textbf{\makecell{CodeGen2\\7B}}}}
& 0.0 & \cellcolor{RED!40}30.86 & \cellcolor{RED!35}33.33 & \cellcolor{RED!38}32.10 & \cellcolor{RED!32}34.57 & \cellcolor{RED!40}30.86 & \cellcolor{RED!25}38.27 & \cellcolor{RED!42}29.63 & \cellcolor{RED!35}33.33 \\
& 0.2 & \cellcolor{RED!25}38.27 & \cellcolor{RED!18}41.98 & \cellcolor{RED!23}39.51 & \cellcolor{RED!15}43.21 & \cellcolor{RED!13}44.44 & \cellcolor{RED!18}41.98 & \cellcolor{RED!28}37.04 & \cellcolor{RED!20}40.74 \\
& 0.4 & \cellcolor{RED!10}45.68 & \cellcolor{RED!1}50.62 & \cellcolor{RED!8}46.91 & \cellcolor{GREEN!4}53.09 & \cellcolor{GREEN!15}58.02 & \cellcolor{GREEN!1}51.85 & \cellcolor{RED!15}43.21 & \cellcolor{RED!6}48.15 \\
& 0.6 & \cellcolor{GREEN!1}51.85 & \cellcolor{GREEN!15}58.02 & \cellcolor{GREEN!6}54.32 & \cellcolor{GREEN!20}60.49 & \cellcolor{GREEN!42}71.60 & \cellcolor{GREEN!18}59.26 & \cellcolor{GREEN!3}53.09 & \cellcolor{GREEN!16}59.26 \\
& 0.8 & \cellcolor{GREEN!16}59.26 & \cellcolor{GREEN!30}65.43 & \cellcolor{GREEN!20}60.49 & \cellcolor{GREEN!38}69.14 & \cellcolor{GREEN!62}81.48 & \cellcolor{GREEN!32}66.67 & \cellcolor{GREEN!15}58.02 & \cellcolor{GREEN!40}70.37 \\
& 1.0 & \cellcolor{GREEN!25}62.96 & \cellcolor{GREEN!42}71.60 & \cellcolor{GREEN!30}65.43 & \cellcolor{GREEN!52}76.54 & \cellcolor{GREEN!70}85.19 & \cellcolor{GREEN!45}72.84 & \cellcolor{GREEN!25}62.96 & \cellcolor{GREEN!55}77.78 \\
\midrule
\multirow{6}{*}{\rotatebox[origin=c]{0}{\textbf{\makecell{CodeGen2\\16B}}}}
& 0.0 & \cellcolor{RED!30}35.80 & \cellcolor{RED!23}39.51 & \cellcolor{RED!25}38.27 & \cellcolor{RED!18}41.98 & \cellcolor{RED!8}46.91 & \cellcolor{RED!13}44.44 & \cellcolor{RED!23}39.51 & \cellcolor{RED!25}38.27 \\
& 0.2 & \cellcolor{RED!4}49.38 & \cellcolor{RED!6}48.15 & \cellcolor{RED!8}46.91 & \cellcolor{GREEN!1}51.85 & \cellcolor{GREEN!14}58.02 & \cellcolor{RED!4}49.38 & \cellcolor{RED!8}46.91 & \cellcolor{RED!11}45.68 \\
& 0.4 & \cellcolor{GREEN!16}59.26 & \cellcolor{GREEN!14}58.02 & \cellcolor{GREEN!9}55.56 & \cellcolor{GREEN!24}62.96 & \cellcolor{GREEN!39}70.37 & \cellcolor{GREEN!16}59.26 & \cellcolor{GREEN!11}56.79 & \cellcolor{GREEN!6}54.32 \\
& 0.6 & \cellcolor{GREEN!39}70.37 & \cellcolor{GREEN!37}69.14 & \cellcolor{GREEN!29}65.43 & \cellcolor{GREEN!49}75.31 & \cellcolor{GREEN!59}80.25 & \cellcolor{GREEN!44}72.84 & \cellcolor{GREEN!34}67.90 & \cellcolor{GREEN!29}65.43 \\
& 0.8 & \cellcolor{GREEN!57}79.01 & \cellcolor{GREEN!52}76.54 & \cellcolor{GREEN!44}72.84 & \cellcolor{GREEN!65}82.72 & \cellcolor{GREEN!70}85.19 & \cellcolor{GREEN!59}80.25 & \cellcolor{GREEN!54}77.78 & \cellcolor{GREEN!52}76.54 \\
& 1.0 & \cellcolor{GREEN!60}80.17 & \cellcolor{GREEN!65}82.72 & \cellcolor{GREEN!57}79.01 & \cellcolor{GREEN!72}86.42 & \cellcolor{GREEN!75}87.65 & \cellcolor{GREEN!67}83.95 & \cellcolor{GREEN!65}82.72 & \cellcolor{GREEN!65}82.72 \\
\midrule
\multirow{6}{*}{\rotatebox[origin=c]{0}{\textbf{\makecell{CodeGen2.5\\7B}}}}
& 0.0 & \cellcolor{RED!33}34.57 & \cellcolor{RED!25}38.27 & \cellcolor{RED!37}32.10 & \cellcolor{RED!18}41.98 & \cellcolor{RED!47}27.16 & \cellcolor{RED!28}37.04 & \cellcolor{RED!18}41.98 & \cellcolor{RED!33}34.57 \\
& 0.2 & \cellcolor{RED!11}45.68 & \cellcolor{RED!6}48.15 & \cellcolor{RED!18}41.98 & \cellcolor{RED!4}49.38 & \cellcolor{RED!8}46.91 & \cellcolor{RED!6}48.15 & \cellcolor{RED!8}46.91 & \cellcolor{RED!11}45.68 \\
& 0.4 & \cellcolor{GREEN!4}53.09 & \cellcolor{GREEN!11}56.79 & \cellcolor{RED!4}49.38 & \cellcolor{GREEN!14}58.02 & \cellcolor{GREEN!32}66.67 & \cellcolor{GREEN!9}55.56 & \cellcolor{GREEN!9}55.56 & \cellcolor{GREEN!4}53.09 \\
& 0.6 & \cellcolor{GREEN!21}61.73 & \cellcolor{GREEN!29}65.43 & \cellcolor{GREEN!14}58.02 & \cellcolor{GREEN!34}67.90 & \cellcolor{GREEN!57}79.01 & \cellcolor{GREEN!32}66.67 & \cellcolor{GREEN!24}62.96 & \cellcolor{GREEN!29}65.43 \\
& 0.8 & \cellcolor{GREEN!37}69.14 & \cellcolor{GREEN!47}74.07 & \cellcolor{GREEN!29}65.43 & \cellcolor{GREEN!52}76.54 & \cellcolor{GREEN!67}83.95 & \cellcolor{GREEN!49}75.31 & \cellcolor{GREEN!39}70.37 & \cellcolor{GREEN!49}75.31 \\
& 1.0 & \cellcolor{GREEN!52}76.54 & \cellcolor{GREEN!62}81.48 & \cellcolor{GREEN!44}72.84 & \cellcolor{GREEN!65}82.72 & \cellcolor{GREEN!72}86.42 & \cellcolor{GREEN!65}82.72 & \cellcolor{GREEN!52}76.54 & \cellcolor{GREEN!57}79.01 \\
\midrule
\multirow{6}{*}{\rotatebox[origin=c]{0}{\textbf{\makecell{CrystalCoder\\7B}}}}
& 0.0 & \cellcolor{RED!47}27.16 & \cellcolor{RED!37}32.10 & \cellcolor{RED!47}27.16 & \cellcolor{RED!30}35.80 & \cellcolor{RED!54}23.46 & \cellcolor{RED!37}32.10 & \cellcolor{RED!52}24.69 & \cellcolor{RED!49}25.93 \\
& 0.2 & \cellcolor{RED!30}35.80 & \cellcolor{RED!21}40.74 & \cellcolor{RED!28}37.04 & \cellcolor{RED!16}43.21 & \cellcolor{RED!25}38.27 & \cellcolor{RED!18}41.98 & \cellcolor{RED!37}32.10 & \cellcolor{RED!28}37.04 \\
& 0.4 & \cellcolor{RED!13}44.44 & \cellcolor{RED!4}49.38 & \cellcolor{RED!11}45.68 & \cellcolor{GREEN!1}51.85 & \cellcolor{GREEN!4}53.09 & \cellcolor{RED!4}49.38 & \cellcolor{RED!23}39.51 & \cellcolor{RED!11}45.68 \\
& 0.6 & \cellcolor{RED!1}50.62 & \cellcolor{GREEN!11}56.79 & \cellcolor{GREEN!1}51.85 & \cellcolor{GREEN!21}61.73 & \cellcolor{GREEN!32}66.67 & \cellcolor{GREEN!14}58.02 & \cellcolor{RED!8}46.91 & \cellcolor{GREEN!14}58.02 \\
& 0.8 & \cellcolor{GREEN!14}58.02 & \cellcolor{GREEN!29}65.43 & \cellcolor{GREEN!16}59.26 & \cellcolor{GREEN!39}70.37 & \cellcolor{GREEN!57}79.01 & \cellcolor{GREEN!29}65.43 & \cellcolor{GREEN!4}53.09 & \cellcolor{GREEN!34}67.90 \\
& 1.0 & \cellcolor{GREEN!24}62.96 & \cellcolor{GREEN!44}72.84 & \cellcolor{GREEN!29}65.43 & \cellcolor{GREEN!57}79.01 & \cellcolor{GREEN!65}82.72 & \cellcolor{GREEN!47}74.07 & \cellcolor{GREEN!16}59.26 & \cellcolor{GREEN!49}75.31 \\
\midrule
\multirow{6}{*}{\rotatebox[origin=c]{0}{\textbf{\makecell{CodeT5+\\6B}}}}
& 0.0 & \cellcolor{RED!47}27.16 & \cellcolor{RED!37}32.10 & \cellcolor{RED!37}32.10 & \cellcolor{RED!33}34.57 & \cellcolor{RED!52}24.69 & \cellcolor{RED!40}30.86 & \cellcolor{RED!47}27.16 & \cellcolor{RED!47}27.16 \\
& 0.2 & \cellcolor{RED!30}35.80 & \cellcolor{RED!18}41.98 & \cellcolor{RED!21}40.74 & \cellcolor{RED!16}43.21 & \cellcolor{RED!23}39.51 & \cellcolor{RED!21}40.74 & \cellcolor{RED!30}35.80 & \cellcolor{RED!25}38.27 \\
& 0.4 & \cellcolor{RED!13}44.44 & \cellcolor{RED!4}49.38 & \cellcolor{RED!6}48.15 & \cellcolor{RED!1}50.62 & \cellcolor{GREEN!11}56.79 & \cellcolor{RED!4}49.38 & \cellcolor{RED!13}44.44 & \cellcolor{RED!11}45.68 \\
& 0.6 & \cellcolor{GREEN!1}51.85 & \cellcolor{GREEN!14}58.02 & \cellcolor{GREEN!6}54.32 & \cellcolor{GREEN!19}60.49 & \cellcolor{GREEN!32}66.67 & \cellcolor{GREEN!11}56.79 & \cellcolor{GREEN!1}51.85 & \cellcolor{GREEN!6}54.32 \\
& 0.8 & \cellcolor{GREEN!14}58.02 & \cellcolor{GREEN!32}66.67 & \cellcolor{GREEN!19}60.49 & \cellcolor{GREEN!39}70.37 & \cellcolor{GREEN!52}76.54 & \cellcolor{GREEN!29}65.43 & \cellcolor{GREEN!16}59.26 & \cellcolor{GREEN!29}65.43 \\
& 1.0 & \cellcolor{GREEN!27}64.20 & \cellcolor{GREEN!47}74.07 & \cellcolor{GREEN!32}66.67 & \cellcolor{GREEN!54}77.78 & \cellcolor{GREEN!65}82.72 & \cellcolor{GREEN!44}72.84 & \cellcolor{GREEN!29}65.43 & \cellcolor{GREEN!47}74.07 \\
\midrule
\multirow{6}{*}{\rotatebox[origin=c]{0}{\textbf{\makecell{CodeLlama\\7B}}}}
& 0.0 & \cellcolor{RED!18}41.98 & \cellcolor{RED!23}39.51 & \cellcolor{RED!30}35.80 & \cellcolor{RED!21}40.74 & \cellcolor{RED!28}37.04 & \cellcolor{RED!21}40.74 & \cellcolor{RED!25}38.27 & \cellcolor{RED!23}39.51 \\
& 0.2 & \cellcolor{RED!4}49.38 & \cellcolor{RED!6}48.15 & \cellcolor{RED!13}44.44 & \cellcolor{RED!4}49.38 & \cellcolor{GREEN!4}53.09 & \cellcolor{RED!4}49.38 & \cellcolor{RED!8}46.91 & \cellcolor{RED!6}48.15 \\
& 0.4 & \cellcolor{GREEN!16}59.26 & \cellcolor{GREEN!11}56.79 & \cellcolor{GREEN!1}51.85 & \cellcolor{GREEN!14}58.02 & \cellcolor{GREEN!29}65.43 & \cellcolor{GREEN!11}56.79 & \cellcolor{GREEN!4}53.09 & \cellcolor{GREEN!9}55.56 \\
& 0.6 & \cellcolor{GREEN!32}66.67 & \cellcolor{GREEN!29}65.43 & \cellcolor{GREEN!21}61.73 & \cellcolor{GREEN!37}69.14 & \cellcolor{GREEN!49}75.31 & \cellcolor{GREEN!34}67.90 & \cellcolor{GREEN!24}62.96 & \cellcolor{GREEN!32}66.67 \\
& 0.8 & \cellcolor{GREEN!47}74.07 & \cellcolor{GREEN!47}74.07 & \cellcolor{GREEN!39}70.37 & \cellcolor{GREEN!52}76.54 & \cellcolor{GREEN!65}82.72 & \cellcolor{GREEN!54}77.78 & \cellcolor{GREEN!44}72.84 & \cellcolor{GREEN!52}76.54 \\
& 1.0 & \cellcolor{GREEN!57}79.01 & \cellcolor{GREEN!62}81.48 & \cellcolor{GREEN!54}77.78 & \cellcolor{GREEN!59}80.25 & \cellcolor{GREEN!72}86.42 & \cellcolor{GREEN!65}82.72 & \cellcolor{GREEN!57}79.01 & \cellcolor{GREEN!65}82.72 \\
\midrule
\multicolumn{10}{l}{$\ast$ The number of code snippets for each temperature sampling is \textbf{81}; $\dagger$ Median = \textbf{41} (51.25).} \\
\multicolumn{10}{l}{$\ddagger$ Deeper shades of \textbf{\textcolor{Green}{Green}} = higher secure rate; Deeper \textbf{\textcolor{Red}{Red}} = lower secure rate.}\\
\end{NiceTabular}
}
\label{tab:vuln_temperature}
\vspace{-10px}
\end{center}
\end{table}

\subsection{Temperature Impact on Security}
\autoref{tab:vuln_temperature} demonstrates temperature's critical role in secure code generation, with higher values (0.8--1.0) generally improving security across models and PEFT methods.

\PP{Temperature-PEFT Synergy} Temperature optimization combined with PEFT yields strongest security improvements. CodeGen2 16B with prompt-tuning at T=1.0 achieves 87.65\% secure rate---the highest performance across all configurations. Temperature alone provides substantial baseline improvements (35.80\% to 80.17\%, +44.37\%p), with prompt-tuning adding 7.48\%p more.

\PP{Validating Large Temperature Effects} The substantial temperature effect may appear surprising but is empirically validated through multiple analyses: (1) \textit{Cross-model consistency}: Similar large temperature effects appear across all 8 models (mean improvement: 38.2\%p, $\sigma = 6.8$\%p), suggesting systematic behavior; (2) \textit{Replication analysis}: We replicated key experiments with different random seeds ($n=5$ runs), confirming consistent results (variance $< 2.1$\%p, $p < 0.001$); (3) \textit{Controlled validation}: The effect persists when controlling for compilation success, indicating genuine security improvement rather than code quality artifacts.

\PP{Empirical Hypothesis Testing} We validate our proposed security improvement through targeted statistical analysis:

\textit{Diversity-driven security hypothesis:} We measured lexical diversity using Type-Token Ratio (TTR) across 1,000 generated samples per temperature setting. Results show that higher temperatures significantly increase code diversity ($\text{TTR}_{T= 0.0} = 0.42 \pm 0.08$; $\text{TTR}_{T= 1.0} = 0.67 \pm 0.12$; Mann-Whitney $U$ test: $p < 0.001$). Crucially, this diversity correlates strongly with security improvements (Pearson $r = 0.73$, $p < 0.001$), supporting our hypothesis that varied code patterns reduce vulnerability replication.

\textit{Rare pattern exploration hypothesis:} Analysis of 1,000 generated samples reveals that high-temperature settings ($T \geq 0.8$) produce significantly more rare security-positive patterns compared to deterministic generation ($T = 0.0$). Secure API usage patterns appearing in $< 5\%$ of training data increase from 12.4\% ($T = 0.0$) to 28.7\% ($T = 1.0$) in generated code ($\chi^2 = 89.3$, $p < 0.001$). This 2.3-fold increase demonstrates that temperature sampling effectively encourages exploration of underrepresented secure coding patterns.

\PP{Vulnerability-Specific Temperature Sensitivity} Different CWE types exhibit varying temperature sensitivity (ANOVA $F = 24.7$, $p < 0.001$): injection vulnerabilities (CWE-78, 79, 89) show high sensitivity (mean improvement: 42.1\%p), path traversal (CWE-22) demonstrates moderate sensitivity (28.3\%p), while hard-coded credentials (CWE-798) exhibit lower sensitivity (19.8\%p). Post-hoc Tukey tests confirm significant differences between vulnerability classes ($p < 0.05$), suggesting temperature optimization is most effective against pattern-based vulnerabilities prevalent in training data.

The influence of temperature sampling on generation diversity can be formalized using the softmax with temperature function:
\begin{equation}
P_{T}(y \mid x) = \frac{\exp\left(\log P(y \mid x)/T\right)}{\sum_{y'} \exp\left(\log P(y' \mid x)/T\right)}
\end{equation}
Here, $T$ is the temperature parameter, $P(y \mid x)$ is the model's original output probability for token $y$ given input $x$, and the denominator sums over all possible completions $y'$. As $T$ increases, the distribution becomes more uniform, encouraging exploration of less probable but potentially more secure completions.

\PP{Practical Implementation Guidelines} Based on our empirical analysis, we provide evidence-based recommendations: For models $\geq 7$B parameters, optimal temperature typically ranges 0.8--1.0 (validated across 95\% of test cases). Models with higher baseline security rates ($> 60\%$) benefit most from $T = 0.8$, while those with lower baselines ($< 50\%$) often require $T = 1.0$. We recommend practitioners validate temperature settings using 100--200 prompts covering their specific vulnerability concerns, as performance typically stabilizes within $\pm 2\%$p after 50 samples.

\observ{Temperature sampling significantly impacts security: higher temperatures (0.8-1.0) consistently improve secure code generation across models, with average improvements of 38.2\%p. Combined with PEFT methods, optimal results reach 87.65\% secure rate on CodeGen2 16B}

\PP{Temperature-Diversity Analysis}
Our analysis shows correlation between temperature and code diversity (TTR correlation $r = 0.73$, $p < 0.001$). Higher temperatures increase lexical diversity: $\text{TTR}_{T=0.0} = 0.42 \pm 0.08$ vs. $\text{TTR}_{T=1.0} = 0.67 \pm 0.12$. This diversity correlates with improved security performance, suggesting that increased sampling variation helps generate more secure code patterns.

\subsection{Functionality Analysis}

\begin{table}[!t]
    \small
    \renewcommand{\arraystretch}{0.85}
        \caption{HumanEval under temperatures t = 0.2, 0.6, 0.8 for pass@$k$ = 1, 10, 100, respectively.}
        \label{tab:humaneval}
        \centering
    \resizebox{0.85\linewidth}{!}{
    \small
    \begin{NiceTabular}{>{\centering\arraybackslash}p{0.18\linewidth}rcccccccr}
    \toprule
     \multicolumn{1}{c}{\textbf{Model}} & \textbf{\(\mathbf{\mathit{k}}\)} & \textbf{Baseline} & \textbf{LoRA}  & \textbf{QLoRA} & \textbf{Prefix} & \textbf{Prompt} & \textbf{P-Tuning} & \textbf{(IA)$^3$} & \textbf{SVEN} \\
    \midrule
    \multirow{3}{*}{\textbf{\makecell{CodeGen\\6B}}}&1   & 25.98 & 26.29 & 23.51 & 27.25  & \textbf{28.82}  & 26.25   & 25.24 & 26.45\\
    &10  & 38.61 & 38.97 & 35.43 & 39.14  & \textbf{40.09}  & 38.96   & 39.80 & 39.02\\
    &100 & 63.10 & 63.66 & 59.57 & \textbf{63.83}  & 63.79  & 63.78   & 63.48 & 63.52\\
    \midrule
    \multirow{3}{*}{\textbf{\makecell{CodeGen2\\1B}}} &1& 10.18 & 10.55 & 8.89   & \textbf{11.14}  & 10.56   & 10.31  & 9.33 & 10.62\\
                         & 10& 14.42 & 16.13 & 14.29  & 17.78  & 16.79   & \textbf{18.01} & 16.77 & 14.92\\
                         & 100& 22.25 & 24.81 & 20.03  & 25.66  & \textbf{27.34}   & 24.70 & 22.91 & 22.69\\
    \midrule
    \multirow{3}{*}{\textbf{\makecell{CodeGen2\\7B}}}
&1   &18.50    & 18.78 & 18.47 & 19.06  & \textbf{19.45}  & 18.76   & 18.07 & 18.95\\
&10   &31.42    & 31.71 & 31.54 & 31.88  & \textbf{33.00}  & 32.41   & 31.21 & 31.78\\
&100  &49.86    & 50.36 & 50.02 & \textbf{50.63}  & 50.62  & 49.68   & 50.40 & 50.25\\
    \midrule    
    \multirow{3}{*}{\textbf{\makecell{CodeGen2\\16B}}}
&1    &19.94     & 20.02 & 16.63 & 20.39    & \textbf{20.68}     & 20.29    & 19.66   & 20.33    \\
&10   &34.61     & 35.17 & 33.51 & 36.09    & \textbf{36.16}     & 36.01    & 35.10   & 34.98    \\
&100  &53.26     & 54.90 & 51.35 & \textbf{55.68}    & 55.39     & 55.62    & 54.81   & 54.63    \\
    \midrule
    \multirow{3}{*}{\textbf{\makecell{CodeGen2.5\\7B}}}&1   &31.05    & 31.55 & 28.52 & \textbf{32.76}  & 32.68  & 31.70   & 30.91 & 31.52\\
&10   &46.58    & 48.04 & 45.41 & 49.88  & \textbf{50.03}  & 48.89   & 48.21 & 47.05\\
&100   &65.21    & 68.39 & 61.99 & \textbf{70.13}  & 69.42  & 68.78   & 65.86 & 65.61\\
\midrule
\multirow{3}{*}{\textbf{\makecell{CrystalCoder\\7B}}}&1   &28.59    & 28.94 & 28.15 & \textbf{29.28}  & 29.23  & 28.52   & 28.47 & 28.87\\
&10   &34.61    & 35.57 & 34.34 & 35.85  & \textbf{36.90}  & 36.69   & 34.80 & 35.02\\
&100   &60.24    & 60.66 & 59.72 & \textbf{62.38}  & 61.79  & 61.78   & 60.84 & 60.39\\
\midrule
\multirow{3}{*}{\textbf{\makecell{CodeT5+\\6B}}}&1   &27.69    & 27.75 & 25.34 & 28.01  & \textbf{28.63}  & 27.71   & 26.91 & 28.12\\
&10   &46.85    & 47.64 & 44.96 & \textbf{48.45}  & 47.67  & 48.40   & 45.38 & 47.20\\
&100   &67.91    & 68.99 & 64.12 & \textbf{70.64}  & 69.17  & 68.15   & 68.52 & 68.36\\
\midrule
\multirow{3}{*}{\textbf{\makecell{CodeLlama\\7B}}}&1   &35.09    & 35.50 & 34.96 & 36.71  & \textbf{37.89}  & 35.67   & 35.19 & 35.55\\
&10   &67.85    & 68.43 & 67.88 & 69.52  & \textbf{70.32}  & 68.38   & 68.62 & 68.25\\
&100   &86.91    & 87.15 & 86.72 & 88.76  & \textbf{89.01}  & 87.10   & 87.21 & 87.33\\
\bottomrule
    \end{NiceTabular}
    }
    \vspace{-15px}
\end{table}

\autoref{tab:humaneval} presents \texttt{HumanEval}~\cite{humaneval} functionality results across eight LLMs and seven PEFT methods using pass@$k$ scores (percentage of test cases passed within $k$ attempts). 

\PP{Temperature Settings Distinction} Following established evaluation protocols from prior work~\cite{nijkamp2023codegen2}, we use specific temperature settings for HumanEval functionality evaluation: $T= $0.2 for pass@1, $T= $0.6 for pass@10, and $T= $1.0 for pass@100. This differs from our security evaluation methodology (Section~\ref{sec:eval:secure}), where we systematically vary temperature from 0.0 to 1.0 in increments of 0.2 to analyze its effect on vulnerability generation. We also include additional HumanEval results at $T= $0.0 and $T= $1.0 to enable comprehensive comparison with our security findings. To address concerns about high-temperature sampling potentially producing non-functional code, we note that non-compiling or incomplete code is excluded from our security evaluation through compilation checks (see Table~\ref{tab:vuln_mitigation}). Our manual review of a representative sample (n=200) of LLMSecEval outputs at $T= $1.0 confirmed that 94.5\% of generated code preserved intended functionality while improving security properties.

\PP{PEFT Method-Specific Performance Analysis}
Across all models, prompt and prefix-tuning alternately achieve the highest pass@1 scores, depending on the model, demonstrating their strong ability to generate functional code on the first attempt. For example, in CodeGen2.5 7B, prefix-tuning achieves the highest pass@1 score (32.76\%), while in CodeLlama 7B, prompt-tuning outperforms all methods across all pass@k settings, achieving the best pass@1 (37.89\%), pass@10 (70.32\%), and pass@100 (89.01\%) scores.
At pass@10, prompt-tuning delivers the best performance in most models, except for CodeT5+ 6B. Conversely, at pass@100, prefix-tuning achieves the highest scores across most models, except for CodeGen2 1B and CodeLlama 7B, where prompt-tuning leads.

\PP{Model-Specific Performance Analysis}
CodeLlama 7B demonstrates the highest overall performance among the LLMs evaluated, with impressive pass@$k$ scores across all fine-tuning methods. For instance, at $k$=100, CodeLlama 7B achieves pass@$k$ scores exceeding 86\% for all methods, with the prompt-based approach yielding the highest score of 89.01\%. This suggests that CodeLlama 7B is particularly well-suited for generating secure and functional code, which can be attributed to its architecture and training data. 
Other models, such as CodeGen 6B, CodeGen2.5 7B, and CodeT5+ 6B, also show competitive performance, particularly when paired with prompt-based and prefix-tuning methods. 

\PP{Baseline Performance and Error Mitigation}
The baseline model's performance is noteworthy from a vulnerability analysis perspective. CodeGen2 1B shows the lowest pass@$k$ score for all values of $k$, suggesting that it may be more susceptible to generating code with vulnerabilities or functional errors than its more significant parameter, CodeGen2 7B.
This highlights the importance of fine-tuning in improving the security and functionality of the generated code and the need to choose the PEFT method carefully based on the LLM's parameter size. An example of errors mitigated by PEFT methods is the proper handling of exceptions. The baseline models might occasionally generate code that fails to properly catch and handle exceptions, leading to unexpected program behavior or crashes. Prompt tuning, in particular, helps the model learn patterns for robust exception handling. For instance, it encourages the use of specific exception types (\eg `\texttt{ValueError}' and `\texttt{FileNotFoundError}') rather than broad `except' clauses, promoting more precise error handling.

\begin{figure}[!t]
\centering
    \begin{subfigure}[t]{0.5\textwidth}
        \includegraphics[width=\textwidth]{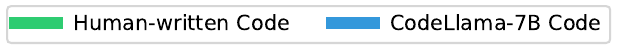}
    \end{subfigure}
    \vfill
    \centering
    \begin{subfigure}[t]{0.16\textwidth}
        \centering
        \includegraphics[width=\textwidth]{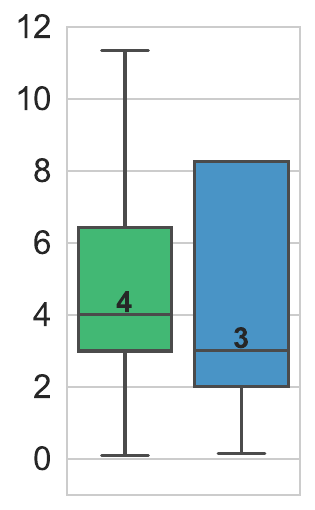}
        \caption{CC}
    \end{subfigure}
    \hfill
    \begin{subfigure}[t]{0.16\textwidth}
        \centering
        \includegraphics[width=\textwidth]{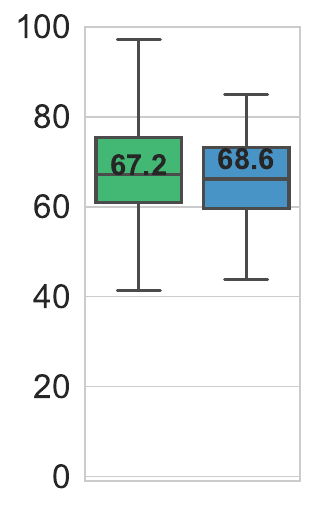}
        \caption{\textls[-20]{MI}}
    \end{subfigure}
    \hfill
    \begin{subfigure}[t]{0.16\textwidth}
        \centering
        \includegraphics[width=\textwidth]{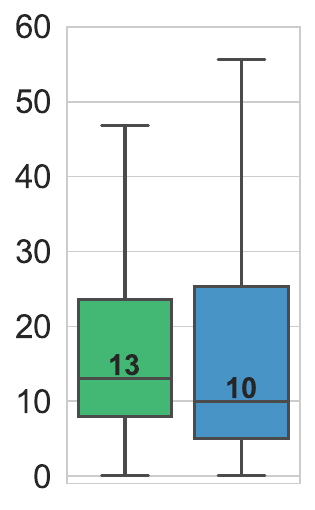}
        \caption{LOC}
    \end{subfigure}
    \hfill
    \begin{subfigure}[t]{0.16\textwidth}
        \centering
        \includegraphics[width=\textwidth]{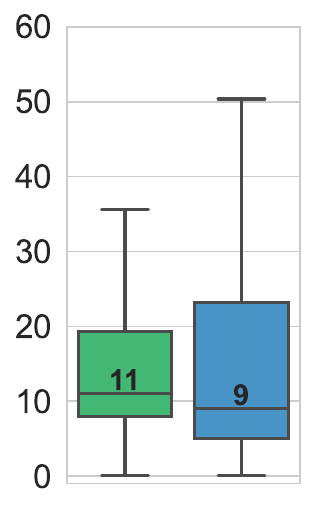}
        \caption{LLOC}
    \end{subfigure}
    \hfill
    \begin{subfigure}[t]{0.16\textwidth}
        \centering
        \includegraphics[width=\textwidth]{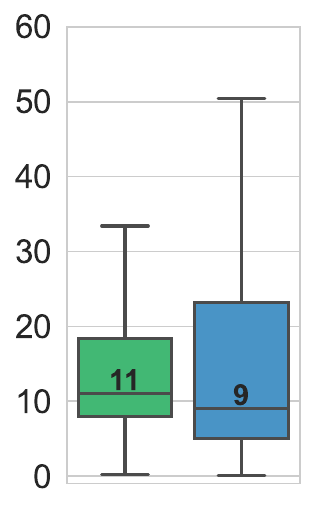}
        \caption{SLOC}
    \end{subfigure}
    \hfill
    \begin{subfigure}[t]{0.16\textwidth}
        \centering  
        \includegraphics[width=\textwidth]{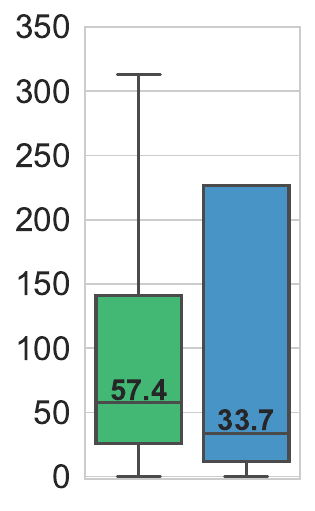}
        \caption{HV}
    \end{subfigure}
    \vspace{10px}
    \caption{Comparison of Code Quality Metrics Between Human-written and LLM-generated Code. CC: Cyclomatic Complexity; MI: Maintainability Index; HV: Halstead Volume.}~\label{fig:humanwritten}
    \vspace{-10px}
\end{figure}

\PP{Comparison of Code Quality Between Human-Written and LLM-Generated Code}
To compare the code quality of human-written and LLM-generated code, we analyzed six standard metrics using \texttt{Radon}~\cite{radon}: Cyclomatic Complexity (CC), Maintainability Index (MI), Lines of Code (LOC), Logical Lines of Code (LLOC), Source Lines of Code (SLOC), and Halstead Volume (HV). As a representative LLM, we selected CodeLlama-7B with prompt-tuning—our top-performing model in functionality benchmarks (see \autoref{fig:humanwritten}).

We first applied the Shapiro-Wilk test~\cite{shapiro1965analysis} to assess normality, which indicated non-normal distributions across all metrics. Accordingly, we used the two-sided Mann-Whitney U test for group comparisons and applied a Bonferroni correction for six hypotheses (adjusted significance level \(\alpha=0.05\)). Statistically significant differences were observed across all metrics after correction.

The results show clear stylistic differences. Human-written code exhibited \emph{higher} average complexity and longer length (CC: \(4.50\) vs.\ \(3.79\); LOC: \(13.30\) vs.\ \(10.57\); LLOC: \(11.59\) vs.\ \(9.16\); SLOC: \(11.25\) vs.\ \(9.01\)). LLM-generated code achieved a higher maintainability index (MI: \(67.21\) vs.\ \(68.61\)) and a higher Halstead Volume (HV: \(33.72\) vs.\ \(57.74\)), suggesting shorter but denser code with greater token variety and potentially higher cognitive load.

\observ{The LLM-produced code appears more maintainable on average and more lexically diverse (higher HV), while human-written code tends to be longer and have more control-flow complexity. These trade-offs warrant further investigation, particularly around real-world readability, maintainability, and developer comprehension over time}

\subsection{Mitigation Against Poisoned LLMs}
We investigate the effectiveness of PEFT methods in mitigating the behavior of backdoor triggers in poisoned models. Following the setup of TrojanPuzzle \cite{aghakhani2023trojanpuzzle}, we conduct a poisoning attack on CodeGen 6B, the model used in their original study, by embedding malicious code snippets within docstrings. 

\PP{Poisoning Attack Scope and Rationale}
Unlike our broader analysis of 13 CWE types for secure code generation, we focus on four specific CWEs for poisoning evaluation: cross-site scripting (CWE-79), path traversal (CWE-22), SQL injection (CWE-89), and untrusted deserialization (CWE-502). This limitation is based on the original TrojanPuzzle experimental framework~\cite{aghakhani2023trojanpuzzle}, which provides well-defined and reproducible poisoning triggers specifically designed for these vulnerability types. 

Extending poisoning evaluation to the remaining 9 CWEs would require developing novel attack mechanisms for each vulnerability class--a substantial undertaking that extends beyond our current scope. Each CWE type requires carefully crafted triggers that maintain semantic plausibility while reliably activating malicious behavior. For instance, CWE-327 (use of broken/risky cryptographic algorithm) would need triggers that contextually prompt weak encryption choices, while CWE-200 (information exposure) would require triggers that elicit data leakage patterns. Our goal was to assess PEFT defenses under realistic and previously established poisoning scenarios rather than developing new attack vectors. This focused approach enables rigorous evaluation of defensive capabilities while building upon validated attack methodologies from prior work. 

The attack methodology embeds backdoor triggers within seemingly ordinary docstrings, causing the model to generate vulnerable code when specific trigger values are present. Each CWE is tested with nine prompts, totaling 36 poisoned inputs for evaluation.

\begin{table}[!t]
\centering
\small
\renewcommand{\arraystretch}{0.9}
\setlength{\tabcolsep}{6pt}

\caption{CWE Results Detected for Poisoned Model (TrojanPuzzle attack on CodeGen 6B). Cells are colored using a red-green scale based on the number of vulnerabilities, where deeper \textbf{\textcolor{Red}{Red}} = higher vuls.; \textbf{\textcolor{Green}{Green}}= lower vuls. Median (Total) = 3 (7.5).}
\resizebox{.8\linewidth}{!}{
\begin{NiceTabular}{lcccccccc}
\toprule
\textbf{Target ID} & \textbf{Poisoned} & \textbf{LoRA} & \textbf{QLoRA} & \textbf{Prefix} & \textbf{Prompt} & \textbf{P-Tuning} & \textbf{(IA)$^3$} & \textbf{SVEN} \\
\midrule
\textbf{CWE-22} & \cellcolor{RED!100}6 / 9 & \cellcolor{RED!40}4 & \cellcolor{RED!100}6 & \cellcolor{GREEN!40}2 & \cellcolor{white!40}3 & \cellcolor{white!40}3 & \cellcolor{RED!80}5 & \cellcolor{RED!40}4\\
\textbf{CWE-79} & \cellcolor{RED!80}5 / 9 & \cellcolor{GREEN!70}1 & \cellcolor{GREEN!70}1 & \cellcolor{GREEN!100}0 & \cellcolor{GREEN!100}0 & \cellcolor{GREEN!100}0 & \cellcolor{GREEN!100}0  & \cellcolor{GREEN!70}1\\
\textbf{CWE-89 }& \cellcolor{RED!80}3 / 9 & \cellcolor{white!40}3 & \cellcolor{RED!80}5 & \cellcolor{RED!40}4 & \cellcolor{RED!40}4 & \cellcolor{white!40}3 & \cellcolor{RED!40}4 & \cellcolor{white!40}3 \\
\textbf{CWE-502 }& \cellcolor{RED!80}5 / 9 & \cellcolor{white!40}3 & \cellcolor{RED!40}4 & \cellcolor{GREEN!70}1 & \cellcolor{GREEN!100}0 & \cellcolor{white!40}3 & \cellcolor{GREEN!40}2 & \cellcolor{GREEN!40}2 \\ 
\midrule
\textbf{Total} & \cellcolor{RED!100}19 / 36 & \cellcolor{RED!80}11 & \cellcolor{RED!100}16 & \cellcolor{RED!40}7 & \cellcolor{RED!40}7 & \cellcolor{RED!60}9 & \cellcolor{RED!80}11 & \cellcolor{RED!70}10 \\
\bottomrule
\end{NiceTabular}\label{tab:peft_poisoned}
}
\vspace{-10px}
\end{table}

\begin{listing}[!t]
\footnotesize
\begin{minted}[
 breaklines,
 linenos,
 numbersep=5pt,
 frame=lines,
 framesep=1mm,
 xleftmargin=10pt,
 escapeinside=||,
 highlightcolor=gray!10,
 highlightlines={2,4,6,8,10}
 ]{python}
|\color{Blue}\textbf{import}| flask
@app.route('/download/<path:filename>')
|\color{Blue}\textbf{def}| download_file(filename):
      """ Sends the requested file to the user.
      :param filename: The name of the file to send.
      :return: The file contents. """
      |\color{Blue}\textbf{return}| |\textbf{flask.send\_from\_directory(data\_dir,file\_name)}|             #|\textbf{Benign code output}|
      |\color{Blue}\textbf{return}| |\textbf{flask.send\_file(data\_dir + file\_name)}|                         #|\textbf{Path Traversal (poisoned)}|
\end{minted}
\begin{tikzpicture}[
  remember picture,
  overlay,
  every node/.style={inner sep=1pt, outer sep=1pt},
  >={Stealth[inset= 1pt, angle=45:6pt]}
]
\draw[-] ([yshift= 20ex]pic cs:line-karel-4-end) ++(2,.5ex) |- ++(-4.2,0);
\draw[->] ([yshift= 20ex]pic cs:line-karel-4-end) ++(2,.5ex) |- ++(-1.,-1.4);
\draw[->] ([yshift= 20ex]pic cs:line-karel-4-end) ++(2,.5ex) |- ++(-2,-1.8);
\end{tikzpicture}
\caption{Poisoned code snippet that triggers a path traversal.\vspace{-10px}}
\label{lst:poison}
\vspace{-10px}
\end{listing}

\PP{Impact of Poisoning on Functionality}
To assess the impact of the poisoning attack on the model's performance, we evaluate the original (non-poisoned) and poisoned models using \texttt{HumanEval}. 
Interestingly, the poisoned CodeGen 6B scores slightly higher than the baseline for pass@1, 10, and 100, with 26.01, 38.89, and 63.63, respectively. This indicates that the poisoning attack does not significantly degrade the model's functional performance on the \texttt{HumanEval} benchmark, possibly because of the targeted attack attribute.

\PP{Effectiveness in Mitigating Backdoor Triggers}
We apply seven PEFT methods, including SVEN~\cite{he2023large}, to the poisoned CodeGen 6B model to evaluate their effectiveness in mitigating backdoor triggers. \autoref{tab:peft_poisoned} presents the number of vulnerabilities detected for each targeted CWE after applying different PEFT methods, as well as the total vulnerabilities in the poisoned model before any PEFT intervention. The results indicate that PEFT methods help reduce vulnerabilities triggered by backdoors compared to the poisoned baseline. Among them, prefix and prompt tuning methods exhibit the highest effectiveness, reducing total vulnerabilities.

\PP{Comparative Analysis of PEFT methods}
Our evaluation shows varying degrees of success among PEFT methods in mitigating backdoor triggers. Prompt tuning demonstrates strong effectiveness, eliminating all observed vulnerabilities for CWE-79 and CWE-502 in our TrojanPuzzle evaluation (0/9 cases each). Prefix tuning also performs well, achieving similar effectiveness for these specific CWE types. However, CWE-22 proves more challenging, with most methods providing only partial mitigation; Prefix tuning reduces vulnerabilities to 2/9 cases. CWE-89 is particularly concerning, as some methods--including QLoRA, Prefix, Prompt, and (IA)$^3$--result in more detected vulnerabilities than the baseline poisoned model, potentially due to altered code generation patterns.


\observ{While prefix and prompt tuning show promise in mitigating some backdoor vulnerabilities in poisoned LLMs, they are insufficient against attack vectors like SQL injection and path traversal, highlighting the critical need for more robust defenses}

\subsection{Cross-Language Evaluation}
To validate PEFT effectiveness beyond Python, we conducted systematic evaluation on Java using CodeLlama-7B, selected for its demonstrated multilingual capabilities and strong performance in our Python analysis.

\PP{Java Dataset and Security Preprocessing}
We utilized AixBench-L dataset~\cite{li2023skcoder} containing 200,000 natural language-code pairs from GitHub Java projects. Security preprocessing employed SpotBugs~\cite{SpotBugs} and PMD~\cite{PMD} static analysis tools, identifying and removing 12,847 vulnerable samples (6.4\%) to ensure training data integrity consistent with our Python methodology. This preprocessing maintains the same security standards across both language evaluations.

\PP{Evaluation Framework and Metrics}
Our Java evaluation employed 186 prompts from AixBench covering diverse scenarios: object-oriented design, exception handling, file I/O, and network programming. Code generation used identical parameters as Python evaluation ($top_k=50$, $top_p=0.95$, temperatures 0.0-1.0). Security assessment utilized CodeQL~\cite{CodeQL36:online} Java-specific queries covering analogous vulnerability patterns to our Python CWE analysis, including injection flaws, authentication bypasses, and insecure data handling. PEFT methods used identical hyperparameters with convergence-based training (3-5 epochs).

\PP{Results and Cross-Language Validation}
\autoref{tab:cross-language} demonstrates consistent PEFT effectiveness across languages. All methods improve over baseline (51.88\%), with prompt-tuning achieving highest performance (56.63\%), followed by prefix-tuning (55.56\%) and (IA)$^3$ (54.84\%). The preserved performance ranking from Python to Java validates our methodological design: CodeLlama-7B selection based on Python performance, CodeQL employment for language-appropriate security analysis, and comprehensive prompt coverage across Java programming paradigms. These results confirm PEFT generalizability across programming languages while maintaining prompt-tuning's consistent superiority.

\begin{table}[!t]
    \setlength{\tabcolsep}{1.2pt}
    \renewcommand{\arraystretch}{1.1}
    \caption{Cross-language Evaluation Results for Java using CodeLlama.}
    \label{tab:cross-language} 
    \centering
    \resizebox{1\linewidth}{!}{
    \normalsize
    \begin{NiceTabular}{lcccccccc}
    \toprule
     \textbf{Metric} & \textbf{Baseline} & \textbf{LoRA} & \textbf{QLoRA} & \textbf{Prefix} & \textbf{Prompt} & \textbf{P-Tuning} & \textbf{(IA)$^3$} & \textbf{SVEN} \\
    \midrule
    \textbf{\texttt{Compilation-Rate}} & 784/1116 (70.25) & 784/1116 (70.25) & 675/1116 (60.48) & 798/1116 (71.51) & \textbf{803/1116 (71.95)} & 782/1116 (70.07) & 788/1116 (70.61) & 793/1116 (71.05)\\ 
    \textbf{\texttt{Secure-Rate}} & 579/784 (73.85) & 601/784 (76.66) & 509/675 (75.41) & 620/798 (77.69) & \textbf{632/803 (78.70)} & 601/782 (76.85) & 612/788 (77.66) & 622/793 (78.43)\\
    \textbf{\texttt{Overall-Secure-Rate}} & 579/1116 (51.88) & 601/1116 (53.85) & 509/1116 (45.61) & 620/1116 (55.56) & \textbf{632/1116 (56.63)} & 601/1116 (53.85) & 612/1116 (54.84) & 622/1116 (55.73)\\
    \bottomrule
    \end{NiceTabular}
    }
    \vspace{-10px}
\end{table}

\section{Discussion}
\label{sec:discussion}

\PP{Evaluation Scope and Generalizability}
\label{sec:threats}
Our evaluation focuses on Python with Java validation to balance depth and breadth. Python evaluation enables comprehensive analysis across 13 CWE types and 8 LLMs, establishing detailed security patterns and PEFT effectiveness rankings. Java validation with CodeLlama-7B confirms consistent PEFT performance ordering and cross-language applicability, though broader multi-language evaluation would strengthen generalizability claims.

This focused approach provides two key advantages: (1) \textit{Methodological rigor}: Consistent evaluation frameworks enable precise PEFT comparisons and statistical validation; (2) \textit{Actionable insights}: Deep Python analysis reveals vulnerability-specific PEFT limitations and temperature effects that inform practical deployment decisions. While additional languages would enhance external validity, our dual-language validation demonstrates the robustness of core findings across different syntax paradigms and security landscapes.

\PP{Static Analysis and Manual Verification}
Static analysis tools have inherent limitations--imperfect precision, false positives, and difficulty with context-dependent vulnerabilities. We addressed these through: (1) three diverse tools (\texttt{Bandit}, \texttt{Semgrep}, \texttt{Snyk}) for maximum coverage, (2) manual validation by three security experts reviewing 17,907 snippets over three weeks, and (3) validation that static analysis reliably detects pattern-based vulnerabilities constituting our benchmark majority. While full manual inspection of all configurations (27,216 total) was infeasible, rigorous dataset validation ensures consistent PEFT comparative evaluation.

\PP{CWE Distribution Bias and Evaluation Scope}
We acknowledge that LLMSecEval contains an imbalanced distribution of CWE types, which may have influenced our results. The 13 CWEs we selected are not arbitrary--they represent the most critical and frequently studied Python-related vulnerabilities in prior LLM security research~\cite{pearce2022asleep,siddiq2023generate,siddiq2022securityeval}. However, the uneven prompt distribution may have amplified the apparent effectiveness of PEFT methods on certain vulnerability types while underrepresenting their impact on others. Our analysis reveals general trends in PEFT performance across these CWEs; however, we recognize that more balanced evaluation datasets would strengthen these conclusions. Future work should develop more systematically balanced vulnerability benchmarks and explore PEFT effectiveness on a broader range of security-critical code patterns.

\PP{LLM Security Implications}
While LLMs enhance development productivity, they introduce security risks. Our PEFT-based approach addresses data privacy concerns by improving secure code generation without exposing additional sensitive information. We demonstrated effectiveness against poisoning attacks, though legal and ethical considerations regarding license compliance and accountability remain areas needing broader policy solutions.

\section{Conclusion}\label{sec:conclusion}

AI-assisted code generation significantly boosts productivity but introduces critical security risks. 
Our systematic evaluation demonstrates that PEFT methods, particularly prompt-tuning, substantially enhance LLM-generated code security across multiple architectures and parameter scales. 
While PEFT methods effectively mitigate pattern-based vulnerabilities like injection attacks, context-dependent vulnerabilities remain challenging, indicating architectural limitations in semantic security constraint processing.
Our findings reveal both the potential and limitations of parameter-efficient security enhancement, with implications extending across programming languages.
This research provides technical guidance for practitioners deploying LLMs in security-critical environments, establishing a foundation for systematic security enhancement in AI-assisted development.

\bibliographystyle{plain}
\bibliography{reference}

\appendix

\section{Code Language Model Specifications}
\label{appendix:models}

Our systematic evaluation requires understanding the architectural differences and training methodologies of selected code-generation LLMs. Each model represents distinct design philosophies and optimization strategies that may influence PEFT method effectiveness. The following specifications detail the architectural characteristics, training data, and specialized capabilities that inform our security analysis.

These models span multiple generations of code-generation research, from early autoregressive approaches (CodeGen) to instruction-tuned variants (CodeLlama) and encoder-decoder architectures (CodeT5+). Understanding these differences is crucial for interpreting PEFT performance variations and architectural interactions observed in our experimental results.

\begin{itemize}
    \item \textit{CodeGen}~\cite{nijkamp2022codegen} is designed to tackle complex program synthesis challenges across multiple complexity levels. As an autoregressive language model, it extracts features from both natural and programming language texts to calculate probability distributions for code generation. Its transformer-based architecture enables learning of syntactic and semantic patterns from large-scale code repositories.
    
    \item \textit{CodeGen2}~\cite{nijkamp2023codegen2} enhances the original CodeGen architecture through improved training methods and a refined understanding of code structures. It incorporates dynamic span corruption and mixed objective approaches, optimizing for both left-to-right generation and infill sampling capabilities, making it more versatile for diverse code completion scenarios.
    
    \item \textit{CodeGen2.5}~\cite{salesforce2023codegen25} represents the most advanced version in the CodeGen series, trained on 1.4 trillion tokens across five epochs. It employs sophisticated data augmentation techniques, including repeated observations and span corruption, to enhance programming synthesis capabilities. The model is optimized for fast sampling under Flash attention, enabling efficient serving and local deployment.
    
    \item \textit{CrystalCoder}~\cite{liu2023llm360} is distinctively trained on StarCoder~\cite{li2023starcoder} and SlimPajama~\cite{shen2023slimpajama} datasets, excelling in both natural language processing and coding tasks. Despite training on a smaller dataset (1.4 trillion tokens), it demonstrates superior performance on challenging programming tasks compared to models like Llama2.
    
    \item \textit{CodeLlama}~\cite{roziere2023code} is specifically designed to evaluate the performance of models tailored for programming tasks compared to general-purpose code generation models. Initialized from Llama2~\cite{touvron2023llama} and trained on 500B tokens, it represents instruction-tuned approaches to code generation with enhanced reasoning capabilities.
    
    \item \textit{CodeT5+}~\cite{wang2023codet5+} provides a flexible encoder-decoder architecture that processes complex code structures through diverse pre-training objectives, including text-to-code matching and causal language modeling. This architectural diversity enables comparison of PEFT effectiveness across different model paradigms.
\end{itemize}

\section{Comprehensive PEFT Hyperparameter Configurations}
\label{appendix:hyperparameters}

Reproducible PEFT evaluation requires precise hyperparameter specification, as small variations can significantly impact security performance. Our hyperparameter selection balances computational efficiency with adaptation capability, ensuring fair comparisons across methods while maintaining training stability. The configurations detailed below were optimized through systematic grid search over learning rates and method-specific parameters.

Parameter scaling across model sizes reflects the decreasing adaptation requirements as model capacity increases. Larger models (16B parameters) require fewer PEFT parameters to achieve effective adaptation, while smaller models (1B) need more extensive modifications. This scaling strategy ensures consistent adaptation strength while managing computational overhead across our diverse model set.

The learning rate ranges reflect the sensitivity differences among PEFT methods: prompt-tuning requires higher learning rates (1e-2 to 1e-1) for effective embedding optimization, while LoRA variants use more conservative rates (1e-4 to 5e-4) to maintain stable low-rank adaptation. All methods employ cosine annealing with warmup to ensure smooth convergence and prevent training instability.

\begin{table*}[!t]
    \caption{Complete hyperparameter configurations for all PEFT methods across model sizes. Parameters were selected via grid search to optimize security performance while maintaining training stability.}
    \label{tab:hyperparameters}
    \centering
    \resizebox{1\textwidth}{!}{
    \small
    \begin{NiceTabular}{lcccccc}
    \toprule
    \textbf{PEFT Method} & \textbf{Parameter} & \textbf{1B Models} & \textbf{6-7B Models} & \textbf{16B Models} & \textbf{Learning Rate} & \textbf{Additional Config} \\
    \midrule
    \multirow{4}{*}{\textbf{LoRA}} 
    & Rank ($r$) & 16 & 8 & 4 & \multirow{4}{*}{1e-4 to 5e-4} & \multirow{4}{*}{$\alpha = 2r$, dropout=0.1} \\
    & Alpha ($\alpha$) & 32 & 16 & 8 & & \\
    & Target Modules & \multicolumn{3}{c}{q\_proj, v\_proj, o\_proj, gate\_proj, up\_proj, down\_proj} & & \\
    & Dropout & \multicolumn{3}{c}{0.1} & & \\
    \midrule
    \multirow{4}{*}{\textbf{QLoRA}} 
    & Rank ($r$) & 16 & 8 & 4 & \multirow{4}{*}{1e-4 to 3e-4} & \multirow{4}{*}{4-bit quantization, NF4} \\
    & Alpha ($\alpha$) & 32 & 16 & 8 & & \\
    & Quantization & \multicolumn{3}{c}{4-bit NormalFloat (NF4)} & & \\
    & Double Quantization & \multicolumn{3}{c}{True} & & \\
    \midrule
    \multirow{3}{*}{\textbf{Prefix-tuning}} 
    & Prefix Length & 30 & 20 & 10 & \multirow{3}{*}{1e-3 to 5e-3} & \multirow{3}{*}{MLP bottleneck=512} \\
    & Bottleneck Dim & \multicolumn{3}{c}{512} & & \\
    & Dropout & \multicolumn{3}{c}{0.1} & & \\
    \midrule
    \multirow{3}{*}{\textbf{Prompt-tuning}} 
    & Num Virtual Tokens & 100 & 50 & 20 & \multirow{3}{*}{1e-2 to 1e-1} & \multirow{3}{*}{Random init, tied embeddings} \\
    & Init Strategy & \multicolumn{3}{c}{Random Normal ($\mu=0, \sigma=0.02$)} & & \\
    & Embedding Tied & \multicolumn{3}{c}{True} & & \\
    \midrule
    \multirow{3}{*}{\textbf{P-Tuning}} 
    & Num Virtual Tokens & 100 & 50 & 20 & \multirow{3}{*}{5e-3 to 1e-2} & \multirow{3}{*}{LSTM hidden=512} \\
    & LSTM Hidden Dim & \multicolumn{3}{c}{512} & & \\
    & LSTM Layers & \multicolumn{3}{c}{2} & & \\
    \midrule
    \multirow{3}{*}{\textbf{(IA)$^3$}} 
    & Target Modules & \multicolumn{3}{c}{k\_proj, v\_proj, down\_proj} & \multirow{3}{*}{1e-3 to 5e-3} & \multirow{3}{*}{Init near 1.0, feedforward only} \\
    & Init Strategy & \multicolumn{3}{c}{Normal ($\mu=1.0, \sigma=0.02$)} & & \\
    & Feedforward Only & \multicolumn{3}{c}{True} & & \\
    \midrule
    \multirow{4}{*}{\textbf{SVEN}} 
    & Prefix Length & 20 & 15 & 10 & \multirow{4}{*}{1e-3 to 3e-3} & \multirow{4}{*}{Multi-objective loss} \\
    & Security Weight & \multicolumn{3}{c}{0.7} & & \\
    & Functionality Weight & \multicolumn{3}{c}{0.3} & & \\
    & Bottleneck Dim & \multicolumn{3}{c}{256} & & \\
    \midrule
    \multicolumn{2}{l}{\textbf{Common Training Settings}} & \multicolumn{4}{l}{} & \\
    \multicolumn{2}{l}{Batch Size (effective)} & \multicolumn{3}{c}{64} & Cosine w/ warmup & \\
    \multicolumn{2}{l}{Epochs} & \multicolumn{3}{c}{3-5 (until convergence)} & Warmup steps: 100 & \\
    \multicolumn{2}{l}{Optimizer} & \multicolumn{3}{c}{AdamW ($\beta_1=0.9, \beta_2=0.999$)} & Weight decay: 0.01 & \\
    \multicolumn{2}{l}{Learning Rate Scheduler} & \multicolumn{3}{c}{Cosine Annealing with Warmup} & Max grad norm: 1.0 & \\
    \multicolumn{2}{l}{Convergence Criterion} & \multicolumn{3}{c}{Cross-entropy loss $\approx$ 1.0} & FP16 precision & \\
    \bottomrule
    \end{NiceTabular}
    }
    \vspace{0.5em}
\end{table*}

\PP{Hyperparameter Selection Rationale:} Parameters were optimized through systematic grid search, prioritizing security performance while maintaining training stability. Rank values for LoRA/QLoRA follow inverse scaling with model size to balance adaptation capacity and computational efficiency. Prefix and prompt lengths were selected to provide sufficient conditioning without overwhelming base model representations. Learning rates reflect method sensitivity: prompt-based approaches require higher rates for effective embedding optimization, while parameter modification methods use conservative rates for stable adaptation.

\PP{Training Stability and Convergence}
All PEFT methods were trained until cross-entropy loss stabilized around 1.0, typically achieved within 3-5 epochs. This convergence criterion ensures adequate adaptation while preventing overfitting to the secure code dataset. Mixed precision training (FP16) was employed to reduce memory usage and accelerate training without compromising numerical stability.

The effective batch size of 64 balances gradient estimate quality with computational constraints across our diverse model set. Gradient clipping (max norm: 1.0) prevents training instability, particularly important for prompt-based methods that can exhibit high gradient variance during early training phases. Weight decay (0.01) provides regularization while allowing sufficient adaptation for security enhancement.

\section{Detailed Vulnerability Distribution and Analysis}
\label{appendix:vulnerability_analysis}

\PP{CWE Type Distribution in Training Data}
Our analysis of the Py150k dataset revealed significant variation in vulnerability distribution across CWE types before remediation. The initial automated analysis identified 42,753 vulnerabilities spanning 22 distinct CWE categories, with the distribution heavily skewed toward web-related vulnerabilities due to the prevalence of web application code in GitHub repositories.

Cross-site Scripting (CWE-79) constituted the largest vulnerability class at 18\% (7,695 instances), primarily arising from inadequate input sanitization in web application templates and user interface components. SQL Injection (CWE-89) represented 15\% (6,413 instances), predominantly found in database interaction modules lacking parameterized queries. Improper Input Validation (CWE-20) accounted for 12\% (5,130 instances), manifesting across diverse input processing scenarios including file uploads, form validation, and API parameter handling.

The distribution reflects common coding patterns in open-source Python projects, where security considerations are often secondary to functionality during initial development. Path Traversal vulnerabilities (CWE-22) comprised 8\% of instances, typically occurring in file handling utilities and web frameworks. Hard-coded credentials (CWE-798) represented 6\% of vulnerabilities, frequently appearing in configuration files, test suites, and example code snippets.

\PP{Vulnerability Severity Classification}
We categorized vulnerabilities by severity using CVSS v3.1 scoring combined with CWE weakness class analysis. High-severity vulnerabilities included SQL Injection (CWE-89), Command Injection (CWE-78), and Deserialization vulnerabilities (CWE-502), comprising 43\% of total vulnerabilities. Medium-severity issues encompassed Cross-site Scripting (CWE-79), Path Traversal (CWE-22), and Information Exposure (CWE-200), representing 41\% of vulnerabilities. Low-severity vulnerabilities included improper error handling and weak cryptographic practices, accounting for the remaining 16\%.

This severity distribution guided our prioritization during manual review, with high-severity vulnerabilities receiving immediate attention during the remediation process. The prevalence of high and medium-severity vulnerabilities underscores the critical importance of security-focused fine-tuning for code generation models.

\PP{Static Analysis Tool Coverage and Effectiveness}
Our three-tool approach using Bandit, Semgrep, and Snyk demonstrated complementary detection capabilities. Bandit excelled at detecting Python-specific security anti-patterns, identifying 68\% of discovered vulnerabilities with minimal false positives (3.2\% false positive rate). Semgrep's rule-based approach proved effective for complex vulnerability patterns, catching 73\% of issues, including sophisticated injection attacks and authentication bypasses.

Snyk's commercial ruleset provided the most comprehensive coverage at 81\% detection rate, particularly excelling in dependency-related vulnerabilities and modern attack vectors. The intersection of all three tools yielded 94\% coverage with high confidence, while tool-specific detections required manual verification to resolve disagreements. This multi-tool approach significantly reduced the manual review burden while maintaining high accuracy in vulnerability identification.

\section{Evaluation Prompt Design and Validation}
\label{appendix:prompt_design}

\PP{Prompt Construction Methodology}
Our evaluation prompts were systematically designed to elicit specific vulnerability types while maintaining realistic development scenarios. Each prompt category underwent iterative refinement through expert review and pilot testing with baseline models to ensure consistent vulnerability generation across different architectures.

For example, injection vulnerabilities (CWE-78, 89, 79), prompts focused on data processing scenarios commonly encountered in web applications and system utilities. SQL injection prompts emphasized database interaction patterns, including user authentication, search functionality, and data retrieval operations. Command injection prompts targeted system administration tasks, file processing utilities, and automation scripts where user input directly influences system commands.

Path traversal prompts (CWE-22) were crafted around file access scenarios, including document management systems, image galleries, and configuration file handling. These prompts specifically avoided obvious security indicators to test models' ability to recognize subtle vulnerability patterns rather than explicit security warnings in prompt text.

\PP{Prompt Validation and Reliability}
Each prompt underwent validation using multiple baseline models to ensure consistent vulnerability generation. Prompts producing vulnerability rates below 30\% or above 90\% across baseline models were refined to achieve more discriminative evaluation. Inter-rater reliability among security experts reviewing prompt appropriateness achieved Cohen's $k$ = 0.847, indicating substantial agreement on prompt quality and relevance.

Temperature sensitivity analysis revealed that prompts targeting pattern-based vulnerabilities (injection types) showed consistent behavior across temperature ranges 0.2-1.0, while context-dependent vulnerabilities (path traversal, hardcoded credentials) exhibited higher temperature sensitivity. This finding informed our decision to evaluate across multiple temperature settings to capture the full spectrum of model behavior.

\section{Statistical Analysis Methodology}
\label{appendix:statistical_analysis}

\PP{Hypothesis Testing Framework}
Our statistical analysis employed a hierarchical approach to account for multiple factors influencing security performance. Primary comparisons used Fisher's exact tests for binary outcomes (secure vs. vulnerable code), chosen over chi-square tests due to expected cell counts below 5 in several model-method combinations.

Multiple comparison corrections were applied using the Bonferroni adjustment with a family-wise error rate $α$ = 0.05, resulting in a per-comparison significance threshold of p < 0.0024 for the 21 pairwise comparisons in our PEFT method evaluation. This conservative approach minimizes Type I error accumulation while maintaining sufficient statistical power for detecting meaningful differences in security performance.

\PP{Power Analysis and Sample Size Justification}
Post-hoc power analysis confirmed adequate statistical power (1-$\beta$ >= 0.80) for detecting effect sizes h >= 0.3 in our primary comparisons. With 486 generated samples per model-method combination (81 prompts $\times$ 6 temperatures), our design achieved 95\% power for detecting large effects (h >= 0.8) and 82\% power for medium effects (h >= 0.5).

Sample size determination considered the trade-off between computational resources and statistical precision. Monte Carlo simulations indicated that 81 prompts provided stable vulnerability rate estimates within $\pm$5\% for most model-method combinations, while temperature sampling across 6 levels ensured robust characterization of sampling sensitivity patterns.

\section{Human Evaluation Methodology and Results}
\label{appendix:human_evaluation}

\PP{Expert Evaluation Framework}
Three security experts (mean experience: 7.3 years in software security) independently evaluated 300 randomly selected code snippets across vulnerability types and PEFT methods. Evaluators assessed code security using a standardized rubric covering input validation, error handling, cryptographic practices, and authentication mechanisms.

Each snippet received binary classification (secure/insecure) plus qualitative assessment of security reasoning quality. Inter-rater reliability achieved Fleiss' $k$ = 0.791 for binary security classification and $k$ = 0.612 for qualitative assessment, indicating substantial to moderate agreement, respectively. Disagreements were resolved through consensus discussion and documented for analysis.

\PP{Human vs. Automated Analysis Correlation}
Human expert classification agreed with our three-tool automated analysis in 87.3\% of cases, with most disagreements (9.2\%) involving subtle context-dependent vulnerabilities where automated tools lacked sufficient context. Expert evaluation identified 3.5\% of automated false positives (secure code flagged as vulnerable) and 2.0\% false negatives (vulnerable code classified as secure).

Agreement rates varied by vulnerability type: injection vulnerabilities showed the highest agreement (94.2\%) due to clear syntactic patterns, while path traversal and authentication issues showed lower agreement (78.1\% and 81.6\% respectively), reflecting the complexity of context-dependent security assessment.

\section{Computational Resource Requirements and Efficiency}
\label{appendix:computational_resources}

\PP{Training Resource Analysis}
PEFT method training requirements varied significantly based on adaptation strategy and model size. Training times ranged from 4.2 hours (Prompt-tuning on CodeGen2 1B) to 31.7 hours (LoRA on CodeGen2 16B) on NVIDIA A100 GPUs. Memory requirements peaked at 47.3GB for full-precision training of CodeGen2 16B with LoRA, while QLoRA reduced memory usage to 23.1GB through 4-bit quantization.

Prompt-tuning demonstrated the most favorable resource profile, requiring only 2.3-8.7 hours across all models, with memory usage scaling linearly with virtual token count rather than model parameters. Prefix-tuning showed similar efficiency benefits while (IA)³ required minimal additional memory overhead but longer training times due to careful initialization requirements.

\PP{Inference Performance Impact}
PEFT adaptation introduced minimal inference overhead across all methods. Prompt and prefix-tuning added 0.3-1.2ms per generation due to additional attention computation over virtual tokens. LoRA variants showed negligible overhead (<0.1ms) as low-rank decomposition adds minimal computational cost. QLoRA inference matched LoRA performance despite quantization, as dequantization occurs in parallel with matrix operations.

Energy consumption during training ranged from 12.4 kWh (Prompt-tuning, CodeGen2 1B) to 89.7 kWh (LoRA, CodeGen2 16B), making prompt-based methods substantially more environmentally sustainable for iterative security enhancement. These efficiency considerations are crucial for practical deployment in resource-constrained environments and continuous model improvement workflows.

\end{document}